\documentclass[twocolumn,english,final,showpacs,prl,lengthcheck,superscriptaddress,groupedaddress]{revtex4}
\usepackage[T1]{fontenc}
\usepackage{babel}
\usepackage{amsmath}
\usepackage{soul}
\usepackage[unicode=true, pdfusetitle,
 bookmarks=true,bookmarksnumbered=false,bookmarksopen=false,
 breaklinks=false,pdfborder={0 0 1},backref=false,colorlinks=true]
 {hyperref}
\usepackage{amsfonts}
\usepackage[normalem]{ulem}

\usepackage{color}
\usepackage{graphicx}  % needed for figures
\usepackage{epstopdf}
\usepackage{dcolumn}   % needed for some tables
\usepackage{bm}        % for math
\usepackage{amssymb}   % for math
\usepackage{pifont}    % symbols
\usepackage{soul}

\makeatletter

\@ifundefined{textcolor}{}
{%
 \definecolor{BLACK}{gray}{0}
 \definecolor{WHITE}{gray}{1}
 \definecolor{RED}{rgb}{1,0,0}
 \definecolor{GREEN}{rgb}{0,1,0}
 \definecolor{BLUE}{rgb}{0,0,1}
 \definecolor{CYAN}{cmyk}{1,0,0,0}
 \definecolor{MAGENTA}{cmyk}{0,1,0,0}
 \definecolor{YELLOW}{cmyk}{0,0,1,0}
 \definecolor{DBLUE}{cmyk}{1,0,0,0.3}
 }
\begin{document}

%\linenumbers
% The following information is for internal review, please remove them for submission

\widetext
%\leftline{Version 1.6b}
%\leftline{Primary authors: CB}
%\leftline{To be submitted to (PRL)}
%\leftline{Comment to {\tt d0-run2eb-nnn@fnal.gov} by xxx, yyy}
%\centerline{\em D\O\ INTERNAL DOCUMENT -- NOT FOR PUBLIC DISTRIBUTION}

% the following line is for submission, including submission to the arXiv!!
%\hspace{5.2in} \mbox{Fermilab-Pub-04/xxx-E}

\title{Bulk Nuclear Polarization Enhanced at Room-Temperature by Optical Pumping}
% remove these 3 lines before journal submittal.
%\centerline{author list dated 2 August 2011}
% end removal before journal submittal
%
\affiliation{Department of Physics, Technion, Israel Institute of Technology, Haifa, 32000, Israel}
\affiliation{Department of Chemical Physics, Weizmann Institute of Science, Rehovot, 76100, Israel}
%\thanks{These authors have contributed equally to this work.}
\affiliation{Department of Physics, University of California, Berkeley, CA, 94720-7300, USA}
\affiliation{Nuclear Science Division, Lawrence Berkeley National Laboratory, CA, 94720, USA}

\author{Ran Fischer}\thanks{These authors have contributed equally to this work.}\affiliation{Department of Physics, Technion, Israel Institute of Technology, Haifa, 32000, Israel}
\author{Christian O. Bretschneider}\thanks{These authors have contributed equally to this work.}\affiliation{Department of Chemical Physics, Weizmann Institute of Science, Rehovot, 76100, Israel}
\author{Paz London}\affiliation{Department of Physics, Technion, Israel Institute of Technology, Haifa, 32000, Israel}
\author{Dmitry Budker}\affiliation{Department of Physics, University of California, Berkeley, CA, 94720-7300, USA}\affiliation{Nuclear Science Division, Lawrence Berkeley National Laboratory, CA, 94720, USA}
\author{David Gershoni}\affiliation{Department of Physics, Technion, Israel Institute of Technology, Haifa, 32000, Israel}
\author{Lucio Frydman}\email[E-mail address:]{lucio.frydman@weizmann.ac.il}\affiliation{Department of Chemical Physics, Weizmann Institute of Science, Rehovot, 76100, Israel}

      	 	% D0 authors (remove the first 3 lines
                             		% of this file prior to submission, they
                             		% contain a time stamp for the authorlist)
                             		% (includes institutions and visitors)
%\date{\today}

\begin{abstract}

Bulk $^{13}$C polarization can be strongly enhanced in diamond at room-temperature based on the optical pumping of nitrogen-vacancy color centers. This effect was confirmed by irradiating suitably aligned single-crystals at a $\sim$50 mT field promoting anti-crossings between electronic excited-state levels, followed by shuttling of the sample into a custom-built NMR setup and by subsequent $^{13}$C detection. A nuclear polarization of $\sim$ 0.5\%
 - equivalent to the $^{13}$C polarization achievable by thermal polarization at room temperature at fields of $\sim$2000 T - was measured, and its bulk nature determined based on line shape and relaxation measurements. Positive and negative enhanced polarizations were obtained, with a generally complex but predictable dependence on the magnetic field during optical pumping. Owing to its simplicity, this $^{13}$C room-temperature polarizing strategy provides a promising new addition to existing nuclear hyperpolarization techniques.

%An article usually includes an abstract, a concise summary of the work
%covered at length in the main body of the article. It is used for
%secondary publications and for information retrieval purposes.
%For PRL, the rule of thumb is that the abstract should be less than
%8 lines and the text (excluding authors, abstract but including tables,
%figures and references) should be less than 4 pages (leave about 20 lines
%empty on page 4) in two-column format.
%PRL and PRD papers have to have PACS (Phsyics and Astronomy Classification
%Scheme) numbers. Please see {\tt http://www.aip.org/pacs/} for the numbers
%relevant to your paper. A set of standard references can be found at the
%end of this example paper.
\end{abstract}

\maketitle

\begin{figure*}
\centering{}
\begin{minipage}[t]{17.8cm}
\begin{minipage}[c]{56.5mm}
 \begin{center}
\includegraphics[width=53mm,height=50mm]{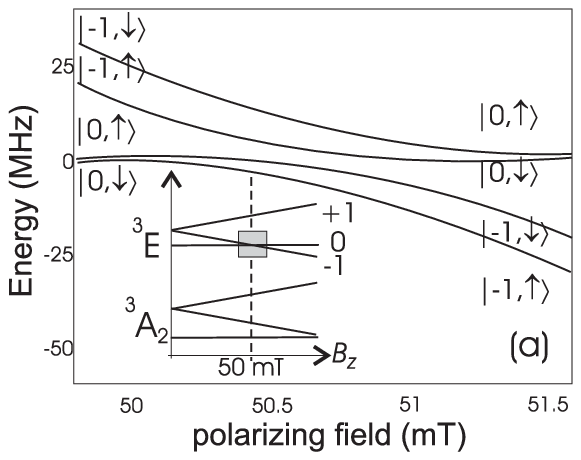} 
\par\end{center}
\end{minipage}
\begin{minipage}[c]{56.5mm}
\begin{center}
\includegraphics[width=53mm,height=50mm]{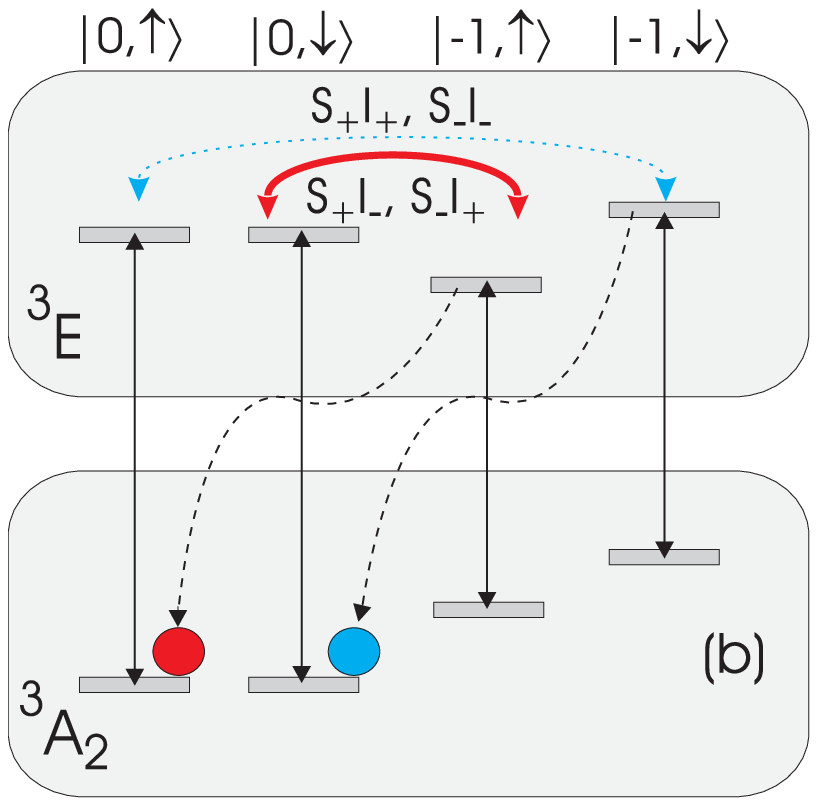} 
\par\end{center}
\end{minipage}
\begin{minipage}[c]{56.5mm}
\begin{center}
\includegraphics[width=53mm,height=50mm]{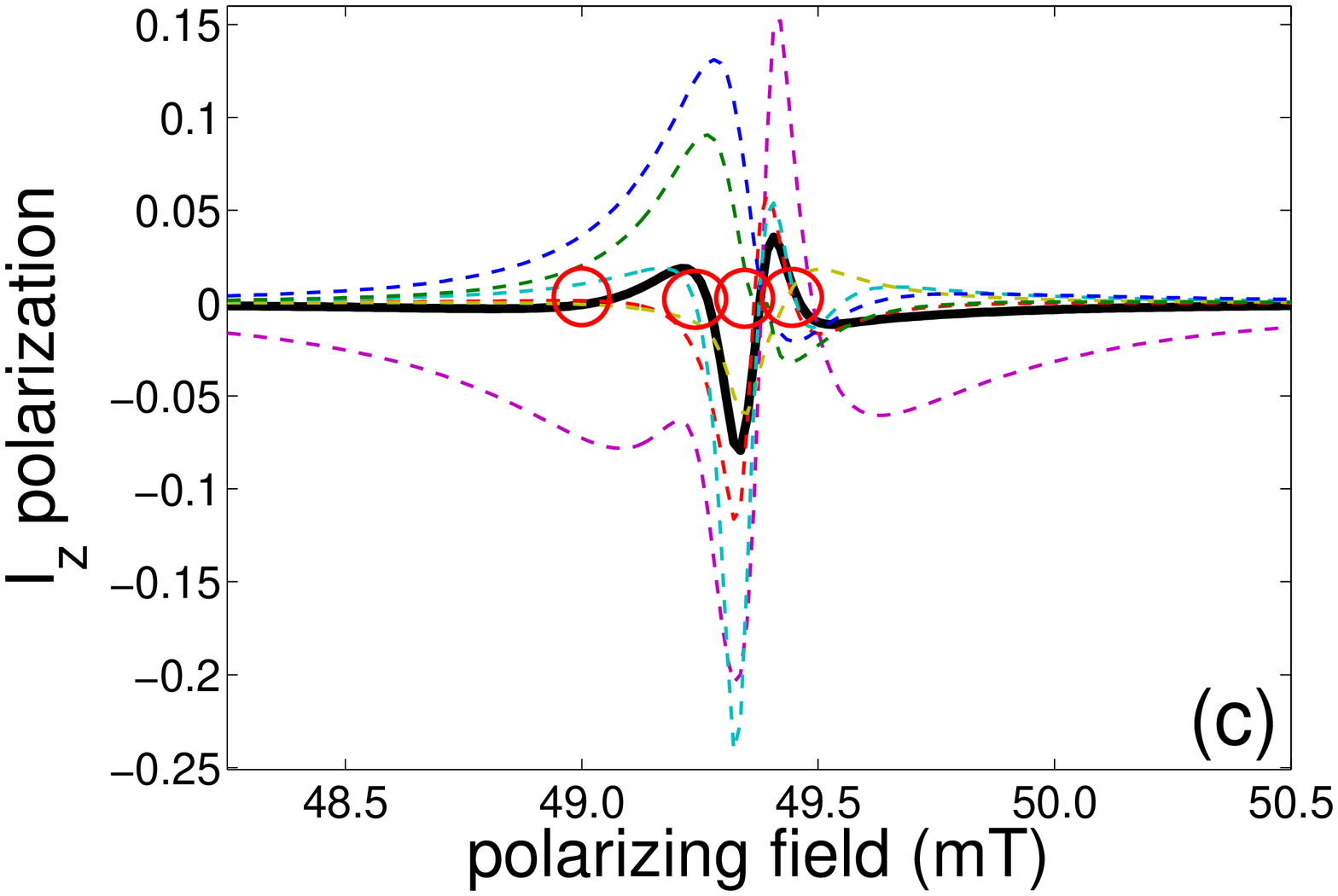} 
\par\end{center}
\end{minipage}
\caption{\label{one} {\small (Color online) (a) Schematic energy diagram of the electronic ground ($^{3}\!A_2$) and excited ($^{3}E$) triplet states of an NV center at room-temperature as a function of an axial magnetic field $B_z$. Numbers denote the electronic spin state, while arrows refer to nuclear spin-${1\over 2}$ $^{13}$C. The magnified sketch shows the region at approximately 50 mT in which level anti-crossings occur in between the excited state sub-levels. 
(b) Simplified schematic description of the polarization transfer processes between the electronic and nuclear sub-states under optical pumping in the presence of the hyperfine interaction. Black solid and dashed arrows represent optical and spin-selective relaxation transitions, and red (cyan) arrows illustrate the mixing of the $|\textit{-}1,\uparrow\rangle$ and $|0,\uparrow\rangle$ ($|\textit{-}1,\downarrow\rangle$ and $|0,\downarrow\rangle$) energy levels, resulting in a positive (negative) nuclear polarization. (c) Nuclear polarizations of $^{13}$C spins as a function of an axial magnetic field as determined by density matrix toy-model simulations. The bold line describes the mean polarization resulting from averaging over six individual orientations of the hyperfine tensor (dashed lines) \cite{Chil06}. Further details on this calculation can be found in the Supplementary Materials \cite{Supp13}. Notice the multiple zero crossings (red circles) of the nuclear polarization due to an effective averaging over sub-ensembles with different hyperfine tensors.}}
\end{minipage}
\end{figure*}

{\sl Introduction.}--- Nuclear Magnetic Resonance (NMR) is commonly used to extract molecular-level information in a wide variety of physical, chemical and biological scenarios. One of NMR's most distinctive characteristics is the low energies it involves. This makes the method remarkably non-invasive but leads to intrinsically low signal-to-noise ratios (SNR), factoring in both small thermal spin polarizations and relatively low frequencies. Over the last decades several strategies have been proposed to bypass these limitations, particularly by generating large, non-thermal spin-polarized states. Most prominent among these techniques is Dynamic Nuclear Polarization (DNP) \cite{Arde03,Wolb04,Golm06,Joo06,Bowe08,Legg10}, where radicals are irradiated in a high-field cryogenic environment, driving their unpaired electron spins away from equilibrium and thereby leading to highly-polarized nuclei on co-mixed, nearby target molecules. This and other forms of hyperpolarization can increase NMR signals by several orders of magnitude, significantly expanding the scope of NMR applications in chemistry and biomedicine \cite{Kurh11}. In the past, free radicals arising in imperfect diamond crystals have also been the target of cryogenic DNP NMR investigations \cite{Reyn98,Casa11}. Those studies have been recently extended to exploit the electronic spin states of negatively-charged Nitrogen-Vacancy (NV) centers in diamond. These color centers have been proposed for a variety of applications including spintronics \cite{Grub97,Epst05,Chil05,Dutt07,Hans08} and ultrasensitive magnetometry \cite{Tayl08,Maze08,Bala08,Acos10}. A particularly appealing aspect of NV-doped diamonds is the simplicity with which they can deliver highly polarized electronic populations by optical pumping at room temperature; several studies have demonstrated that the ensuing electronic polarization can be transferred via hyperfine interactions to nearby nuclei and indirectly detected via the NV defects \cite{Jacq09,Neum10,King10,Cai11,Toga11}. In this study we extend such experiments by direct observations of the nuclear $^{13}$C spins by means of NMR. To execute this polarization transfer, the NV centers are optically pumped by laser irradiation, and the electronic spin polarization was transferred to $^{13}$C spins by suitably orienting the diamond crystal in a magnetic field of $\sim$ 50 mT. This leads to an excited-state level anti-crossing, capable of polarizing $^{13}$C spins exhibiting hyperfine couplings to the NV centers \cite{Jacq09,Schm11,Drea12,Smel09,Fisc13}. To characterize the resulting enhanced nuclear polarization, an experimental setup was built which rapidly transferred the optically pumped NV-doped diamonds to a 4.7 T field. Once at high fields, the $^{13}$C spins are excited using a resonant radiofrequency (RF) pulse, and their nuclear polarization is detected by conventional NMR induction. The intensity of the NMR signal detected after optical pumping exceeded by two-three orders of magnitude the signal observed in the same setup after the sample was subject to full thermal relaxation in a 4.7 T field. It was found that the nuclear spins hyperpolarized by these optical means had the same NMR properties as ``bulk'' $^{13}$C spins \cite{Supp13}. 
These included their relaxation, line width and chemical-shift characteristics, which were found identical to those observed in thermally-polarized high-field NMR experiments on the same sample. This ``bulk-like''
behavior is further substantiated by experimental data suggesting that the majority of the observed hyperpolarized $^{13}$C signal, originates from nuclear spins located beyond 1 nm of the paramagnetic defects. The nuclear polarization also revealed a complex dependence on the magnetic field strength during the optical pumping \cite{WangAr}. This dependence, including NMR signal sign alternations, could be explained by the anisotropic nature of the electron-nuclear hyperfine interaction.
  
{\sl $^{13}$C Polarization Enhancement in Optically Pumped Diamond by Energy Level Anti-Crossings.} --- The negatively charged NV color center in diamond is composed of a substitutional nitrogen atom associated with an adjacent lattice vacancy. This center is characterized by electronic spin-triplets in both the ground ($^{3}\!A_2$) and the excited ($^{3}$E) state. At room temperature these S = 1 states exhibit zero-field splittings between the $m_S$ = 0 and the $m_S$ = $\pm$1 spin sublevels of 2.87 GHz and 1.42 GHz, respectively (Fig. 1a). Optical transitions are primarily spin conserving \cite{Mans06,Dohe11}. However, the presence of a non-radiative, spin-selective relaxation mechanism involving an intersystem crossing with singlet levels, can alter the electronic spin populations. Under suitable optical irradiation the ensuing pumping can be very effective, leading even at room-temperature, to a nearly full population of the $m_s$ = 0 electronic states. This electronic polarization can be transferred to nearby $^{13}$C's in the diamond lattice by suitably tuning the external field (Fig. 1b); transfers to nuclei possessing hyperfine couplings $\ge$ 0.2 MHz have thus been demonstrated \cite{Drea12}. To illustrate the nature of this transfer, we neglect for simplicity the $^{14}$N-related spin interactions, and consider a single NV-$^{13}$C spin Hamiltonian
\begin{figure*}
\centering{}
\begin{minipage}[t]{17.8cm}
\begin{minipage}[c]{90mm}
 \begin{center}
\includegraphics[width=91mm,height=50mm]{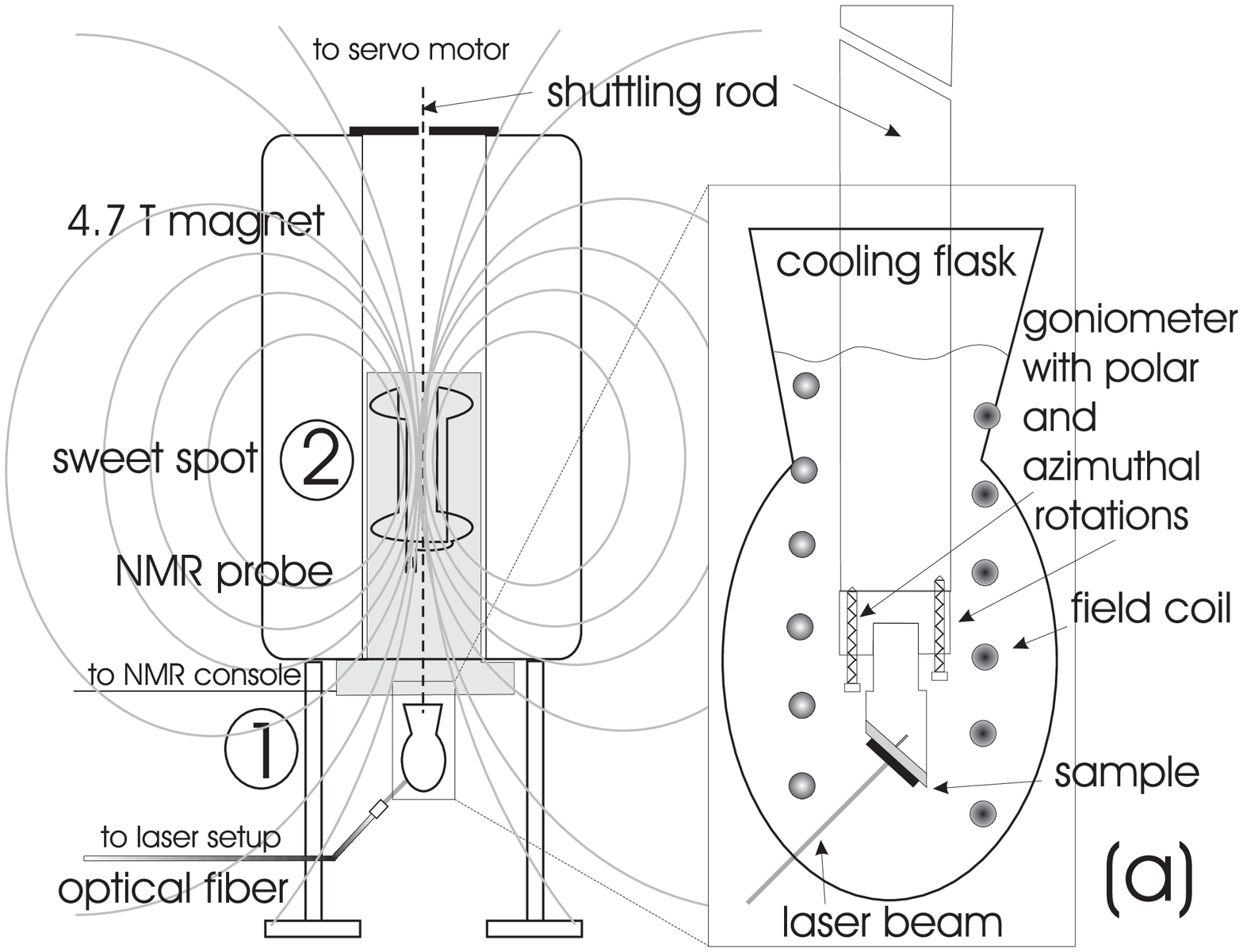} 
\par\end{center}
\end{minipage}
\begin{minipage}[c]{80mm}
\begin{center}
\includegraphics[width=80mm,height=45mm]{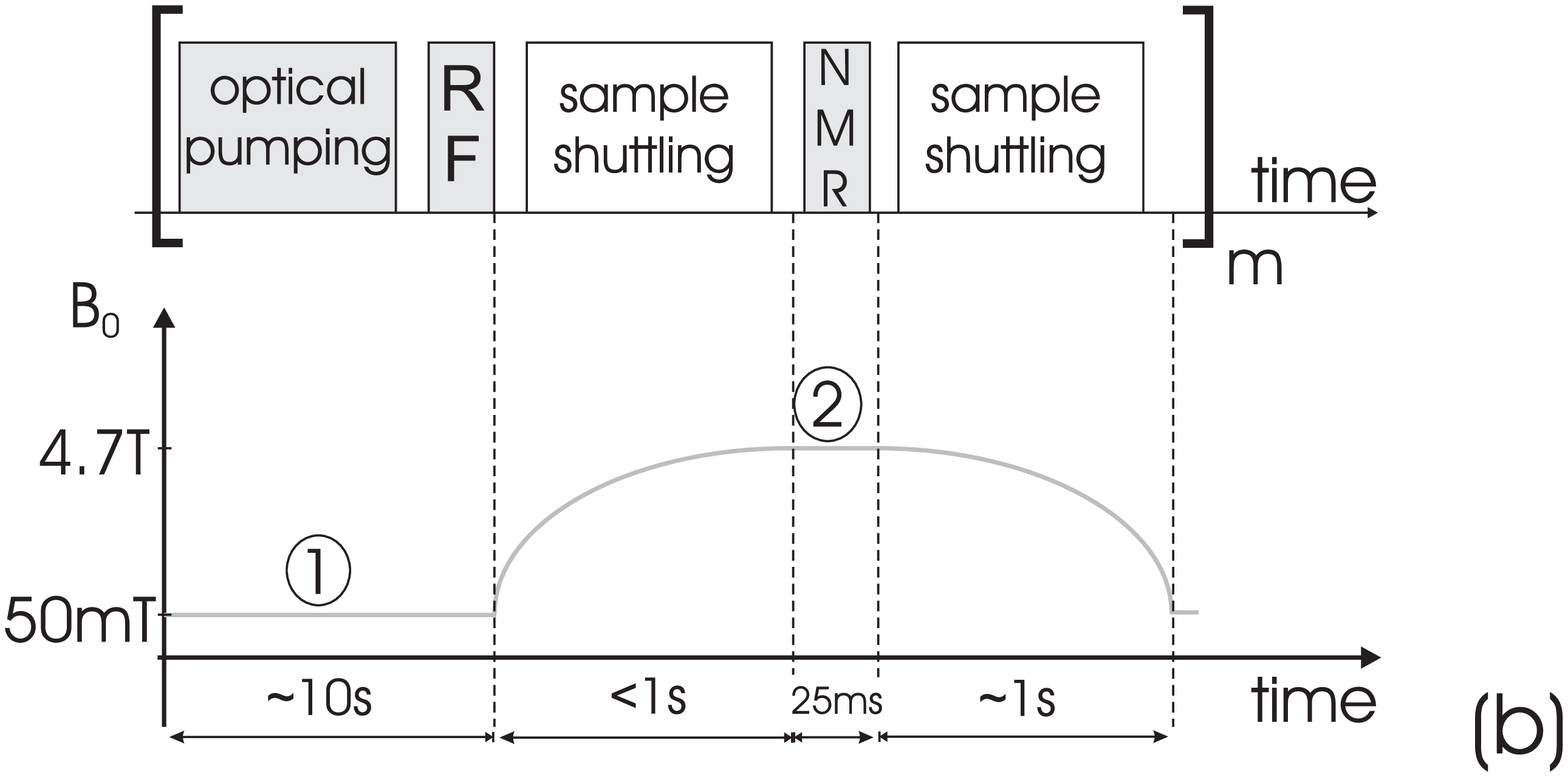} 
\par\end{center}
\end{minipage}
\caption{\label{two} {\small (a-b) Schematic description of the experimental setup and summary of the events occurring during the experiments described in this paper. Experiments begin with laser pumping of the electronic spins in a $\sim$ 50 mT axial field facilitating polarization transfer to the nuclear spins. (Nuclear spins could also be manipulated at this low field by applying resonant RF pulses at a frequency of $\sim$ 545 kHz.) The sample is then shuttled from the polarizing field \ding{192} to the ``sweet spot'' 
of a high-field magnet \ding{193}, where the $^{13}$C spins are subject to a pulsed spin-echo detection. The sample is then returned to the polarizing field for a repeated pumping and further signal averaging. 
The samples used (D01, D02) are natural abundance $^{13}$C single-crystals and exhibit an estimated concentration of NV centers of 3 and 10 ppm, respectively \cite{Supp13}. The pumping laser was connected to the optical setup via a 600 $\mu$m core multimode fiber irradiating an area slightly larger than the diamond sample, with a 532 nm laser beam with an adjustable power of up to 10 W. The setup was used for NMR and ODMR measurements \cite{Supp13}. 
The field was fine-tuned via an ancillary coil. During optical pumping, the sample was immersed in a water-filled flask to avoid excessive heating, and the concomitant distortion of the single-crystal alignment setup. The sample holder was connected by a shuttling rod to a servo motor enabling its transfer (in $<$1s) into and out of high-field NMR setup. The bulk $^{13}$C polarization was monitored in this setup using a custom-built NMR probe enabling free shuttling through a 12 mm Helmholtz coil configuration tuned to 50.55 MHz.
}}
\end{minipage}
\end{figure*}
\begin{equation} \hat{H} = D_{ES} \hat{S}^2_z + (\gamma_{NV}\hat{S} + \gamma_{^{13}C}\hat{I})\cdot\vec{B} + \hat{S} \cdot A_{IS} \cdot \hat{I}  .\end{equation}
Here $\hat{I}$  and $\hat{S}$ are the nuclear and electron spin operators, $D_{ES}$ is the excited-state zero-field splitting, $\gamma_{NV}$ = 2.8$\cdot 10^{4}$ MHz/T and $\gamma_{^{13}C}$ = 10 MHz/T are the corresponding gyromagnetic ratios, and $A_{IS}$ is the hyperfine tensor. It is also assumed that, as was experimentally the case, a positive, axial magnetic field $\vec{B}$ = $B_z\:\hat{z}$ has been aligned with the NV axis, leaving the electronic spin an eigenstate of $\hat{S}_z$. In general, the zero-field and electronic Zeeman splitting dominating Eq. (1), suppresses electronic $\leftrightarrow$ nuclear spin exchanges. Still, as illustrated in Fig. 1(a), energy levels can be tuned with $B_z$: if a field of $\sim$ 50 mT is chosen, the contributions of the first two terms in the electronic Hamiltonian will balance out for the electronic \{$|0\rangle$, $|\textit{-}1\rangle$\} subspace. 
This in turn will lead to an electronic anti-crossing due to the presence of the hyperfine interaction with $^{13}$C nuclei, resulting in a 
transfer between the optically pumped electronic and the nuclear spin populations \{$|\!\!\uparrow\rangle$,$|\!\!\downarrow\rangle$\}. Indeed, when $\displaystyle{||D_{ES}-\gamma_{NV} B_z|\pm \gamma_{^{13}C} B_z| \approx |A_{IS}|}$, the electronic spin is no longer an eigenstate of $\hat{H}$. The system can still be described in a joint manifold of the nuclear and electronic spins, but the eigenstates of $\hat{H}$ under these conditions are mixtures of the $|0,\uparrow\rangle$, $|0,\downarrow\rangle$, $|\textit{-}1,\uparrow\rangle$ and $|\textit{-}1,\downarrow\rangle$ states. If $A_{IS}$ is a so-called isotropic hyperfine interaction possessing solely $\hat{S}_{+}\hat{I}_{-}$, $\hat{S}_{-}\hat{I}_{+}$ ``zero-quantum'' 
terms, only the $|\textit{-}1,\uparrow\rangle$ and $|0,\downarrow\rangle$ states are mixed near the level anti-crossing, leading to an enrichment of the nuclear $|\!\!\uparrow\rangle$ state upon optical pumping (Fig. 1b). By contrast, $^{13}$C's exhibit both isotropic and anisotropic hyperfine characters \cite{Felt09}, resulting in an additional mixing of states involving single- and double-quantum operators $\hat{S}_{z}\hat{I}_{\pm}$, $\hat{S}_{\pm}\hat{I}_{z}$, $\hat{S}_{+}\hat{I}_{+}$ and $\hat{S}_{-}\hat{I}_{-}$. Under optical pumping, these zero- and double quantum terms can drive the nuclei into either polarized or anti-polarized states (Fig. 1b); which one of these states is dominant after the transfer depend on the magnitude of $\vec{B}$ and on the characteristics of $A_{IS}$. This is illustrated by toy-model simulations in Fig. 1(c), which show that multiple sign alterations in the nuclear spin polarization can be expected as a function of the $B_z$ field, when $B_z$ is close to the anti-crossing condition. This calculation is based on solving a master Liouville-von Neumann equation based on the Hamiltonian of Eq. (1). This in turn approximates the polarization build-up of the $^{13}$C nuclear spins via a direct transfer process; its efficiency is determined by the magnitude of the hyperfine interaction, the life-time ($\sim$ 10 ns) of the electronic excited state, and the pumping rate of the NV centers. Such calculations are consistent with observations, showing significant nuclear polarizations (> 10 \%)
facilitated by strong hyperfine interactions (> 0.2 MHz) \cite{Drea12}. In actuality, also weaker hyperfine interactions could enable a low-magnitude polarization transfer throughout the entire diamond sample - which although slow, would become significant once supported by long nuclear $T_1$ relaxation times. In addition, nuclear spin-diffusion events driven by ($\hat{I}_{1+}\hat{I}_{2-}+\hat{I}_{1-}\hat{I}_{2+}$)-like terms originating from the nuclear dipole-dipole interaction, would also assist in transferring the nuclear polarization from strongly-coupled $^{13}$C's towards the bulk $^{13}$C spin reservoir.

To investigate the existence and extent of these polarization transfer mechanisms, which could enable the pumping of significant bulk nuclear polarization by optical means, we assembled the setup illustrated in Fig. 2(a).
The experiment began with careful positioning and alignment of an NV center-endowed diamond single crystal, in a $B_z$ $\sim$ 50 mT field fulfilling the excited-state anti-crossing conditions \cite{Supp13}. Following laser irradiation, the sample was mechanically shuttled within sub-second timescales from the polarizing field to a detection field of 4.7 T, via a servo motor controlled by an NMR console. 
This motion positioned the diamond crystal within a $^{13}$C-tuned Helmholtz-coil circuit; the experiment concluded with the application of an NMR spin-echo sequence to probe the level of $^{13}$C magnetization. Such pump-shuttle-probe processes could be repeated numerous times for signal averaging.
\begin{figure*}
\centering{}
\begin{minipage}[t]{17.8cm}
\begin{minipage}[c]{85mm}
\begin{center}
\includegraphics[width=82mm,height=50mm]{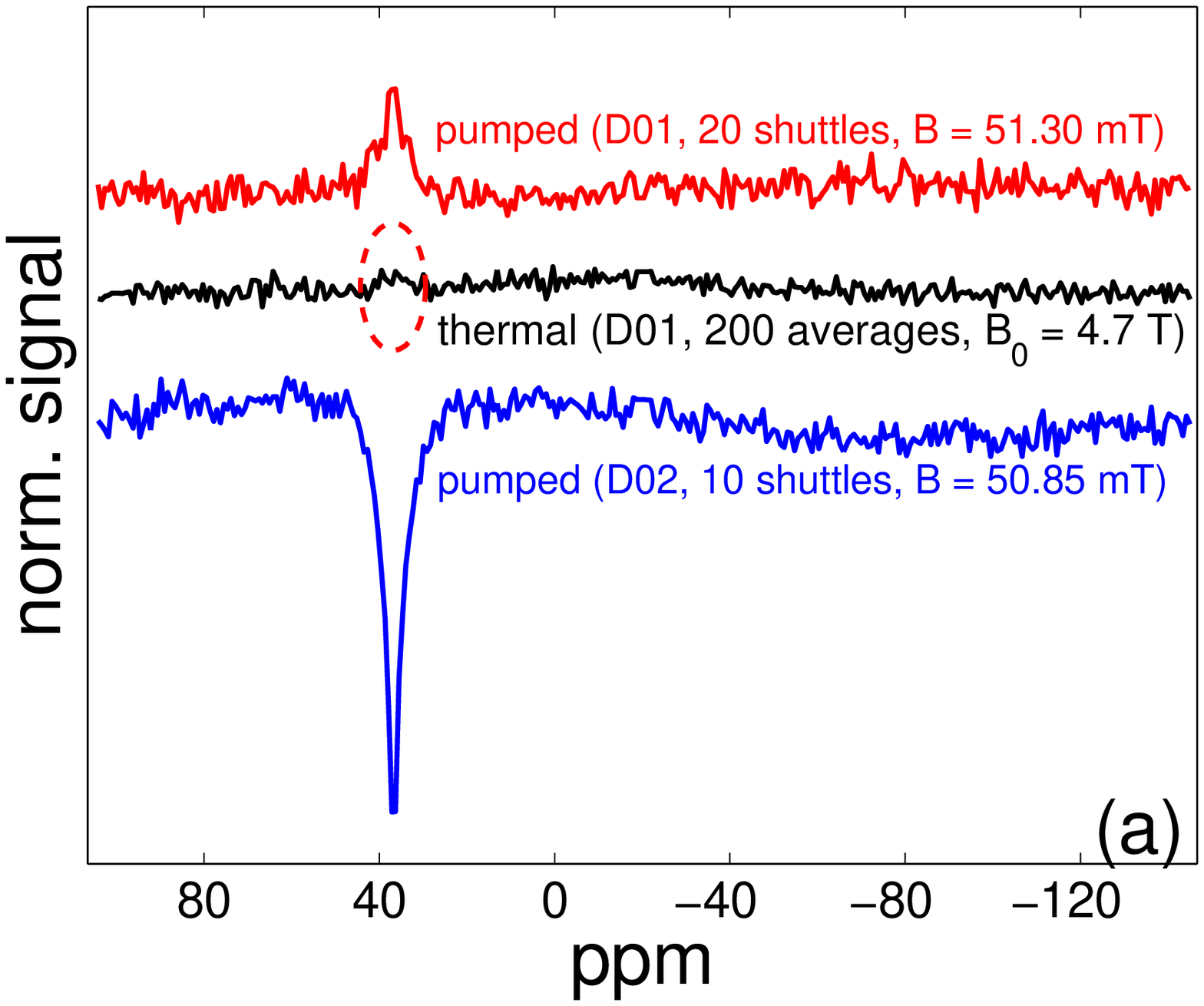} 
\par\end{center}
\end{minipage}
\begin{minipage}[c]{85mm}
\begin{center}
\includegraphics[width=82mm,height=50mm]{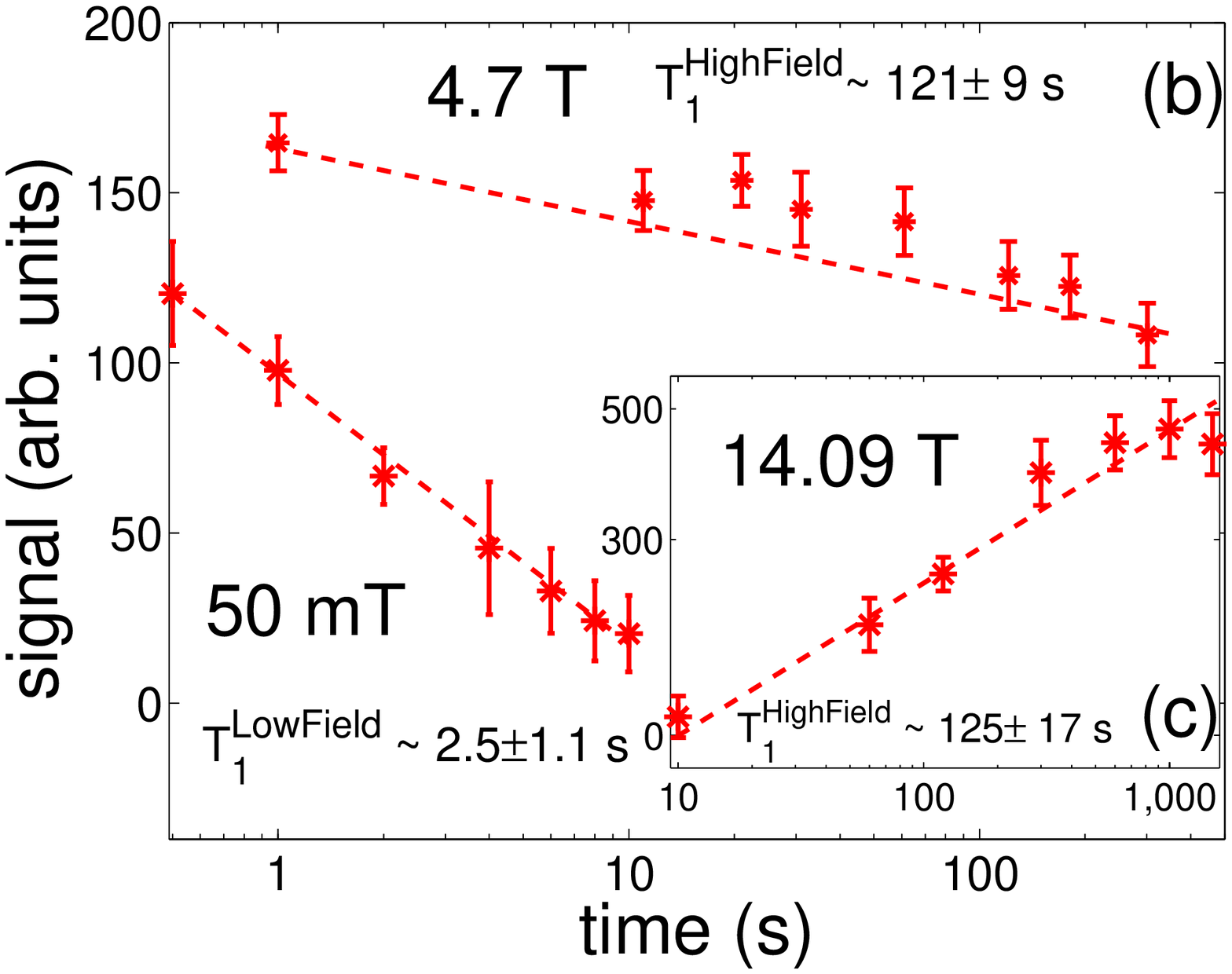} 
\par\end{center}
\end{minipage}

\begin{minipage}[c]{85mm}
\begin{center}
\includegraphics[width=82mm,height=50mm]{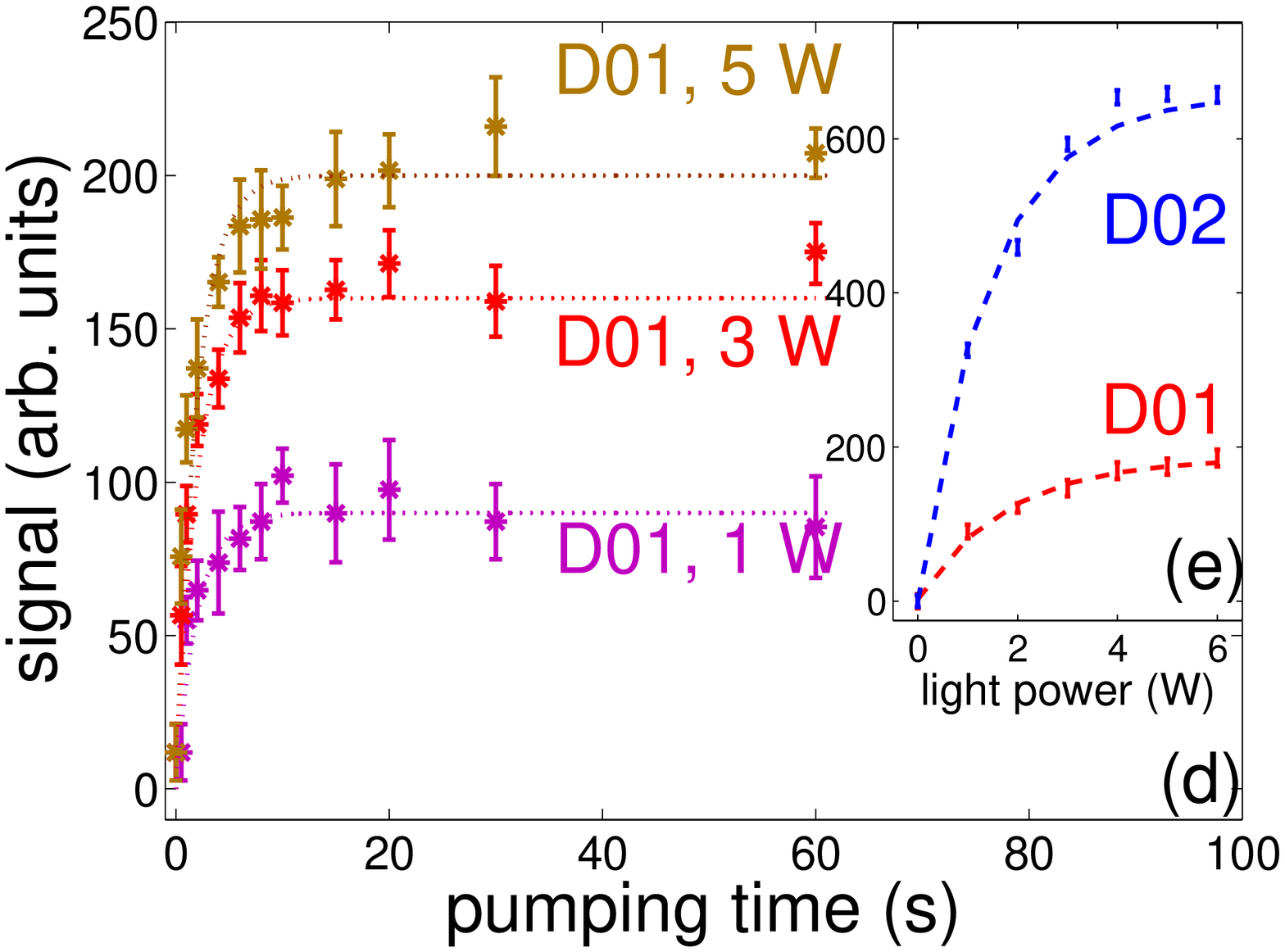} 
\par\end{center}
\end{minipage}
\begin{minipage}[c]{85mm}
\begin{center}
\includegraphics[width=82mm,height=50mm]{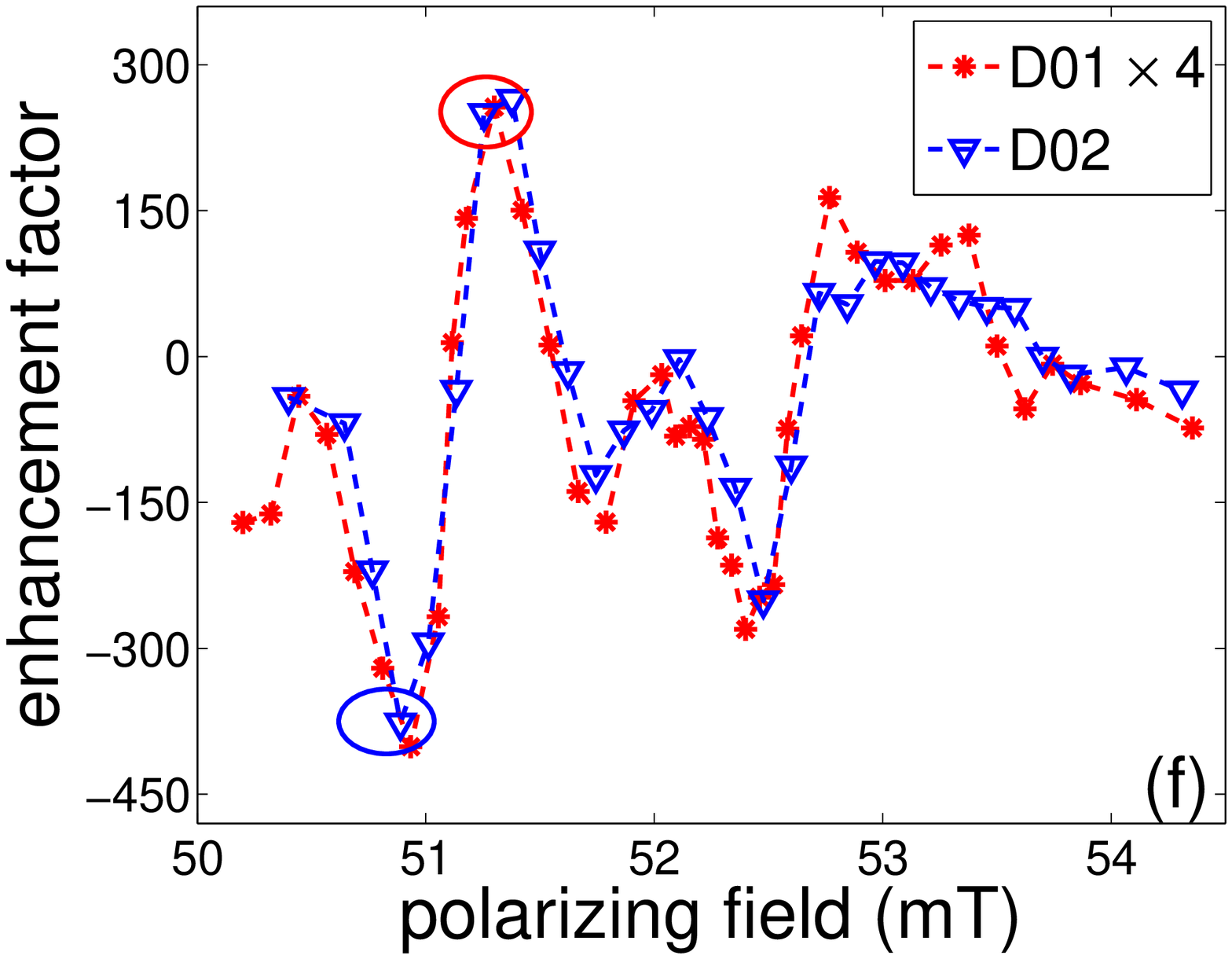} 
\par\end{center}
\end{minipage}
\caption{\label{three} {\small (Color online) Measured experimental data of the diamond samples D01 (red lines, asterisks) and D02 (blue lines, triangles). The dashed lines are guides for the eye. (a) $^{13}$C signal (red and blue line) obtained after 10 s of optical pumping and $^{13}$C signal acquired with signal averaging at 4.7 T for the diamond D01 (black line). The positive and negative polarizations correspond to different polarizing fields and the dashed oval marks the position of the NMR signal, corresponding to the thermal Boltzmann polarization of the sample. 
(b) Spin-lattice relaxation characteristics of the optically pumped signal of the sample D01. The longitudinal relaxation of the $^{13}$C polarization was measured at the two turning points of the shuttling process marked as \ding{192} and \ding{193} in Fig. 2(a), 2(b): the polarizing field, and the observation field (see \cite{Supp13} for similar data from sample D02). (c) Saturation-recovery of bulk $^{13}$C obtained under conventional thermal polarizing conditions on a high-field solid-state spectrometer ($B_0$ = 14.09 T). (d-e) Build-up characteristics of the optically pumped $^{13}$C polarization as a function of the irradiation time for various laser-beam powers and 10 s of pumping. For the D01 sample the signal build-up evolved into a steady-state after $\sim$ 10 s, indicating a limitation of the polarization transfer by the longitudinal relaxation ($T_1^{\textit{\tiny LowField}}$) of the involved nuclei. 
(f) The enhancement of the $^{13}$C SNR as a function of the polarizing field. For both diamond samples identical field dependence was obtained, including positive and negative amplitudes at different polarizing fields. The spectra shown in panel (a) correspond to the encircled points: red circle $B_z$ = 51.30 mT, blue circle $B_z$ = 50.85 mT). Notice that the enhancement of D01 is shown magnified 4$\times$, to stress its similarity to the D02 observable pattern.
}}
\end{minipage}
\end{figure*}

{\sl Results and Discussion.}--- NMR measurements on the samples, D02 and D01, revealed that an average bulk nuclear polarization of $\sim$ 0.50 and 0.12\% 
was achieved by optical pumping of the NV centers.
For D02 the ratio between the optically pumped signal and the signal observed in a thermally polarized ensemble at room-temperature and 4.7 T, the so-called enhancement factor, was determined to be 486 $\pm$ 51 \cite{Supp13}. Taking into account the different polarizing/observation field strengths, the overall enhancement at 50 mT is > 10000 times stronger then the $B_z$-derrived prediction of the thermal Boltzmann polarization. This hyperpolarized $^{13}$C signal showed identical characteristics as signals arising from thermally polarized high-field spin ensembles: within experimental errors, their chemical shift, their spin-lattice relaxation times and linewidths, are all identical (Figs. 3a-c). The results in Fig. 3(d), suggests that the relatively short nuclear relaxation time at 50 mT ultimately limits the amount of electron polarization that can be transferred. 

It follows from these results that the observed hyperpolarized signal is representative of (and therefore probably originates from) the $^{13}$C bulk nuclei, rather than from a subset of nuclei in the immediate vicinity of the NV centers. By contrast, if the $^{13}$C signal were to arise from the immediate surroundings of paramagnetic defects (e.g. $\frac{A_{IS}}{2\pi}\ge$ 10 kHz), shorter relaxation times as well as hyperfine shifts, would be expected driven by the electronic defects \cite{Supp13,Reyn97,Drea13}. The origin of the enhanced nuclear polarization was further investigated by manipulating the nuclear spins with resonant, low-field RF pulses. In particular, a 180$^{\circ}$ pulse of a duration of $t_{\textit{180}}\sim$ 90 $\mu$s was applied at the nominal Larmor frequency of the $^{13}$C spins at the 50 mT polarizing field. Such pulses are resonant solely with the nuclear spin population within its frequency bandwidth ($t^{-1}_{\textit{180}}\sim$ 10 kHz) and will therefore not affect $^{13}$C spins exhibiting strong frequency shifts (e.g. by paramagnetic sources).
The application of such a pulse during the course of low-field optical pumping (in which the majority of the NV centers are found in $m_S$ = 0 state, devoid of hyperfine interaction) should result in an inversion of the entire macroscopic $^{13}$C polarization. By contrast, if the RF pulse is applied after the laser illumination has ended and the electronic population has returned to a thermal Boltzmann distribution, only $^{13}$C spins with hyperfine couplings that are weaker than the Rabi frequency would experience a full inversion. Hence, this experiment distinguishes between nuclear population with hyperfine interaction stronger and weaker than the $\sim$ 10 kHz bandwidth of the used RF pulse. Measuring both signals after reaching a steady-state of the nuclear polarization reveals a full nuclear population flip (Fig. 3 in \cite{Supp13}). This confirms that a majority of the hyperpolarized signal originates from $^{13}$C spins with hyperfine interactions < 10 kHz with the NV centers, corresponding to a minimum distance of $\sim$ 1 nm to electronic paramagnetic defects within the diamond crystal.

From a spin-physics standpoint, it is interesting to note that the relative sign of the $^{13}$C bulk magnetization can be controlled by scanning the polarization field. This stems from the hyperfine interaction which controls the electronic-nuclear spin transfer, as reflected by the simulations in Fig. 1(c). This effect is demonstrated experimentally in Fig. 3(f). The observed dependence of the $^{13}$C polarization on the magnetic field resembles the predictions based on Eq. (1), but exhibits several additional zero-crossings. These sign alternations in the nuclear polarization were identical for both diamond crystals D01 and D02 (Fig. 3f). 
This implies that the observed bulk nuclear polarization sign alterations can be interpreted as manifestations of a general (albeit complex) behavior that the multiple $^{13}$C sub-species coupled by the anisotropic hyperfine interaction with an NV center undergo at each resonance instance \cite{Drea12}. Indeed, the possible positions of $^{13}$C nuclei in the diamond crystal with respect to an NV center, result in an array of individual hyperfine interactions. Therefore, different magnetic field conditions will result for each of these cases, in most distinct patterns for the efficient electronic-nuclear polarization transfers. This occurrence of multiple sign alternations in the nuclear spin polarization contrasts with the magnetic field dependence of $^{15}$N nuclear spin polarizations within the NV center \cite{Jacq09}: $^{15}$N exhibits a nearly isotropic hyperfine interaction with the color center, and thus only a positive nuclear spin polarization is observed. 

{\sl Conclusions.}--- This study shows that in a single-crystal room temperature diamond, a significant $^{13}$C nuclear polarization can be established within seconds, by optical pumping of NV centers at a suitable magnetic field. From this observation, many additional interesting paths emerge to exploit possible synergies between this method of spin polarization and magnetic-resonance experiments. Foremost among these are alternative routes to enhance even further the bulk $^{13}$C polarization; either by manipulating the concentration of NV centers, by $^{13}$C enrichment, or by choosing different irradiation, field or temperature conditions during the pumping. In addition, strategies can be envisioned for transferring the $^{13}$C polarization from the diamond to other molecules, or for using the diamond as a reporter of NMR properties while remaining at low magnetic fields. Interesting opportunities for studies of basic characteristics of spin-spin driven transfers within the diamond as well as to chemical and biological interesting molecules are opened. These and other alternatives to enable a better understanding and a widespread use of this pre-polarizing technique, are being investigated.

% acknowledgement.tex                             
%
The authors are grateful to S. Vega and F. Jelezko for the fruitful discussions. This research was supported by DIP Project 710907 (Ministry of Education and Research, Germany), the EU (through ERC Advanced Grant \#
246754), a Helen and Kimmel Award for Innovative Investigation, the IMOD, the National Science Foundation (D.B.) and the generosity of the Perlman Family Foundation. 
%
   % input acknowledgement

%[4] A. Dréau, J.-R. Maze, M. Lesik, J.F. Roch, and V. Jacques, Phys. Rev. B, 85, 134107 (2012);


\begin{thebibliography}{99}
\bibitem[1]{Arde03} J.H. Ardenkj\ae r-Larsen, B. Fridlund, A.Gram, G. Hansson, M.H. Lerche, R. Servin, M. Thaning, K. Golman, Proc. Nati. Acad. Sci. \textbf{10}, p.10158 (2003);
\bibitem[2]{Wolb04}J. Wolber et al., Nucl. Instrum. Methods Phys. Res., Sect. A, \textbf{526}, 173-181 (2004);
\bibitem[3]{Golm06}K. Golman, R. in't Zandt, M. Thaning, Proc. Natl. Acad. Sci. USA, \textbf{103}, 11270-11275 (2006);
\bibitem[4]{Joo06}C.G. Joo, K.N. Hu, J. Bryant, R. Griffin, J. Am. Chem. Soc., \textbf{128}, 9428-9432 (2006); 
\bibitem[5]{Bowe08}S. Bowen and C. Hilty, Angew. Chem., Int. Ed., \textbf{47}, 5235-5237 (2008);
\bibitem[6]{Legg10}J. Leggett et al., Phys. Chem. Chem. Phys., \textbf{12}, 5883-5892 (2010);
\bibitem[7]{Kurh11} J. Kurhanewicz et al., Neoplasia, \textbf{13}, 81 (2011);
\bibitem[8]{Reyn98} E.C. Reynhardt, G.L. High. J. Chem. Phys, \textbf{10}, Vol. 109, p. 4090 (1998); J. Chem. Phys., \textbf{10}, Vol. 109, 4100 (1998);
\bibitem[9]{Casa11} L.B. Casabianca, A.I. Shames, A.M. Panich, O. Shenderova, L. Frydman, J. Phys. Chem. C, \textbf{115}, 39, pp. 19041-19048, (2011);
\bibitem[10]{Grub97} A. Gruber, A. Dr\"abenstedt, C. Tietz, L. Fleury, J. Wrachtrup, C. von Borczyskowski, Science \textbf{276}, 2012 (1997);
\bibitem[11]{Epst05}R.J. Epstein, F.M. Mendoza, Y.K. Kato, and D.D. Awschalom, Nature Physics \textbf{1}, 94 (2005); 
\bibitem[12]{Chil05}L. Childress, M.V. Gurudev Dutt, J.M. Taylor, A.S. Zibrov, F. Jelezko, J. Wrachtrup, P.R. Hemmer, and M.D. Lukin, Science \textbf{314}, 281 (2006); 
\bibitem[13]{Dutt07}M.V. Gurudev Dutt et al., Science, Vol. 316, p. 1312 (2007);
\bibitem[14]{Hans08}R. Hanson, V. Dobrovitski, A. Feigun, O. Gywat and D. Awschalom Science \textbf{320}, 352 (2008);
\bibitem[15]{Tayl08}J.M. Taylor, P. Cappellaro, L. Childress, L. Jiang, D. Budker, P.R. Hemmer, A. Yacoby, R. Walsworth, M.D. Lukin, Nature Phys. \textbf{4}, 810 (2008).
\bibitem[16]{Maze08}J.R. Maze et al., Nature, \textbf{455}, 644 (2008);
\bibitem[17]{Bala08}G. Balasubramanian et al., Nature, \textbf{455}, 648, (2008);
\bibitem[18]{Acos10}V.M. Acosta, E. Bauch, A. Jarmola, L.J. Zipp, M.P. Ledbetter, D. Budker, Appl. Phys. Lett. \textbf{97}, 174104 (2010);
\bibitem[19]{Jacq09} V. Jacques, P. Neumann, J. Beck, M. Markham, D. Twitchen, J. Meijer, F. Kaiser, G. Balasubramanian, F. Jelezko, J. Wrachtrup, Phys. Rev. Lett., Vol. \textbf{102}, p. 057403 (2009);
\bibitem[20]{Neum10} P. Neumann, J. Beck, M. Steiner, F. Rempp, H. Fedder, P.R. Hemmer, J. Wrachtrup, F. Jelezko, Science, Vol \textbf{329}, p. 542 (2010);
\bibitem[21]{King10}J.P. King, P.J. Coles, J.A. Reimer, Phys. Rev. B, Vol. \textbf{81}, p. 073201, (2010);
\bibitem[22]{Cai11}P. London, J. Scheuer, J.-M. Cai, I. Schwarz, A. Retzker, M.B. Plenio, M. Katagiri, T. Teraji, S. Koizumi, J. Isoya, R. Fischer, L. P. McGuinness, B. Naydenov, F. Jelezko, arXiv:1304.4709.
\bibitem[23]{Toga11}E. Togan, Y. Chu, A. Imamoglu, M.D. Lukin, Nature, Vol. \textbf{478}, p. 497 (2011);
\bibitem[24]{Schm11} B. Smeltzer, L. Childress, A. Gali, New J. Phys. \textbf{13}, 025021 (2011);
\bibitem[25]{Drea12}A. Dr\'{e}au, J.-R. Maze, M. Lesik, J.F. Roch, and V. Jacques, Phys. Rev. B, \textbf{85}, 134107 (2012);
\bibitem[26]{Smel09} B. Smeltzer, J. McIntyre, L. Childress, Phys. Rev. A, Vol. 80, 050302 (2009);
\bibitem[27]{Fisc13} R. Fischer, A. Jarmola, P. Kehayias, D. Budker, Phys. Rev. B, Vol. 87, 125207 (2013);
\bibitem[28]{Supp13} See Supplementary Materials at \textsl{http://link.aps.org/supple-mental/xx.xxxx/PhysRevLett.xxx.xxxxxx} for further information on the alignment process by means of optical detection, density matrix simulations, estimation of the enhancement factor and the average number of spins polarized per NV center.
\bibitem[29]{WangAr} H. Wang, C. Shin1, C. Avalos1, S. Seltzer, D. Budker, A. Pines, V. Bajaj, arXiv:1212.0035
\bibitem[30]{Chil06} L. Childress, Ph.D. Thesis, p. 52, Harvard University (2006);
\bibitem[31]{Mans06} N.B. Manson, J.P. Harrison, M.J. Sellars, Phys. Rev. B, Vol. 74, p. 104303 (2006);
\bibitem[32]{Dohe11}M.W. Doherty, N.B. Manson, P. Delaney and L.C.L. Hollenberg, New Journal of Physics, Vol. 13, p. 025019 (2011);
\bibitem[33]{Felt09} S. Felton, A.M. Edmonds, M.E. Newton, P.M. Martineau, D. Fisher, D.J. Twitchen, J.M. Baker, Phys. Rev. B, Vol 79, p. 075203 (2009);
\bibitem[34]{Reyn97} E.C. Reynhardt, C.J. Terblanche, Chem. Phys. Lett., \textbf{269}, 464-468, (1997);
\bibitem[35]{Drea13} A. Dr\'{e}au, P. Spinicelli, J.R. Maze, J.F. Roch, V. Jacques, Phys. Rev. Lett. 110, 060502 (2013);
\end{thebibliography}
\end{document}

%\linenumbers
% The following information is for internal review, please remove them for submission

\widetext
%\leftline{Version 1.6b}
%\leftline{Primary authors: CB}
%\leftline{To be submitted to (PRL)}
%\leftline{Comment to {\tt d0-run2eb-nnn@fnal.gov} by xxx, yyy}
%\centerline{\em D\O\ INTERNAL DOCUMENT -- NOT FOR PUBLIC DISTRIBUTION}

% the following line is for submission, including submission to the arXiv!!
%\hspace{5.2in} \mbox{Fermilab-Pub-04/xxx-E}

\title{Supplementary Information: Bulk Nuclear Polarization Enhanced at Room-Temperature by Optical Pumping}
% remove these 3 lines before journal submittal.
%\centerline{author list dated 2 August 2011}
% end removal before journal submittal
%
\affiliation{Department of Physics, Technion, Israel Institute of Technology, Haifa, 32000, Israel}
\affiliation{Department of Chemical Physics, Weizmann Institute of Science, Rehovot, 76100, Israel}
\affiliation{Department of Physics, University of California, Berkeley, CA, 94720-7300, USA}
\affiliation{Nuclear Science Division, Lawrence Berkeley National Laboratory, CA, 94720, USA}

\author{Ran Fischer}\affiliation{Department of Physics, Technion, Israel Institute of Technology, Haifa, 32000, Israel}
\author{Christian O. Bretschneider}\affiliation{Department of Chemical Physics, Weizmann Institute of Science, Rehovot, 76100, Israel}
\author{Paz London}\affiliation{Department of Physics, Technion, Israel Institute of Technology, Haifa, 32000, Israel}
\author{Dmitry Budker}\affiliation{Department of Physics, University of California, Berkeley, CA, 94720-7300, USA}\affiliation{Nuclear Science Division, Lawrence Berkeley National Laboratory, CA, 94720, USA}
\author{David Gershoni}\affiliation{Department of Physics, Technion, Israel Institute of Technology, Haifa, 32000, Israel}
\author{Lucio Frydman}\affiliation{Department of Chemical Physics, Weizmann Institute of Science, Rehovot, 76100, Israel}

      	 	% D0 authors (remove the first 3 lines
                             		% of this file prior to submission, they
                             		% contain a time stamp for the authorlist)
                             		% (includes institutions and visitors)
%\date{\today}

%\begin{abstract}
%An article usually includes an abstract, a concise summary of the work
%covered at length in the main body of the article. It is used for
%secondary publications and for information retrieval purposes.
%For PRL, the rule of thumb is that the abstract should be less than
%8 lines and the text (excluding authors, abstract but including tables,
%figures and references) should be less than 4 pages (leave about 20 lines
%empty on page 4) in two-column format.
%PRL and PRD papers have to have PACS (Phsyics and Astronomy Classification
%Scheme) numbers. Please see {\tt http://www.aip.org/pacs/} for the numbers
%relevant to your paper. A set of standard references can be found at the
%end of this example paper.
%\end{abstract}

%\pacs{61.72.jn, 76.60.-k, 78.47.-p, 81.05.ug}
\maketitle
\onecolumngrid

\vspace{-0.5cm}

\section {I. POSITIONING AND ALIGNMENT OF DIAMOND BY OPTICALLY DETECTED MAGNETIC RESONANCE (ODMR)}

Fulfilling the excited-state level anti-crossing (ESLAC) conditions that enable the transfer between electronic and nuclear polarization requires an accurate positioning and alignment of the NV-doped diamond single crystal to one of its crystallographic axes in a $B_z\sim$ 50 mT field. Due to the diamond's tetrahedral lattice structure, the angles between a given crystal axis (e.g. [1,1,1]) and the remaining three axes ([-1,1,1], [1,-1,1], [1,1,-1]) are identical. Hence, it can be assumed that, if an NV center axis is aligned parallel to the external magnetic field $\vec{B}$, the transition frequencies of the corresponding non-aligned NV centers will also be identical. This effect can be exploited during the alignment process of the diamond single-crystal: if three individual ODMR resonance frequencies corresponding to different NV orientations are collapsed into a single line, the magnetic field is aligned to a single orientation and the optical pumping efficiency will be maximized. 
\begin{figure*}[h]
\centering{}
\begin{minipage}[t]{17.8cm}
\begin{minipage}[c]{84.5mm}
\begin{center}
\includegraphics[width=75mm,height=55mm]{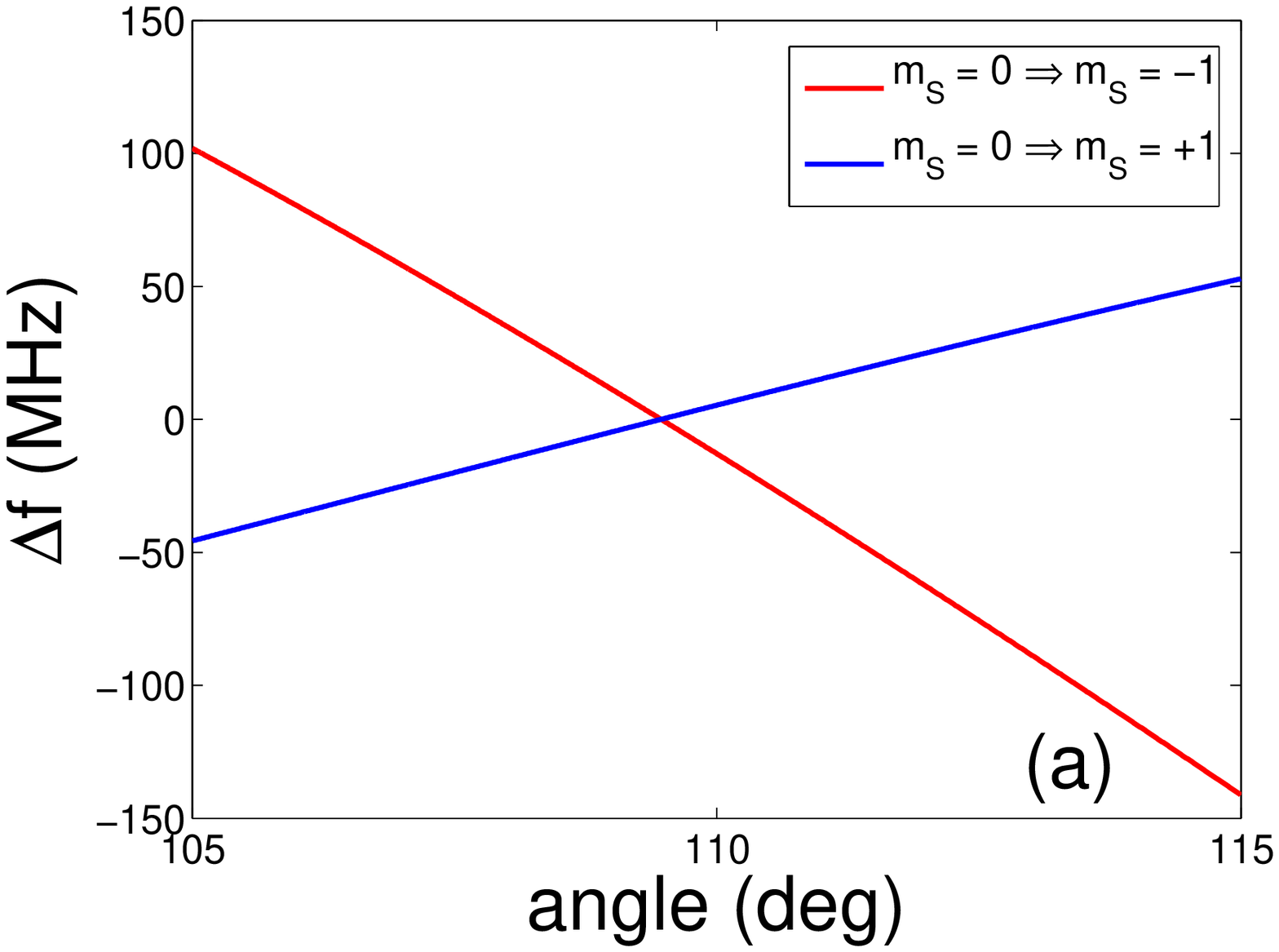} 
\par\end{center}
\end{minipage}
\begin{minipage}[c]{84.5mm}
\begin{center}
\includegraphics[width=75mm,height=55mm]{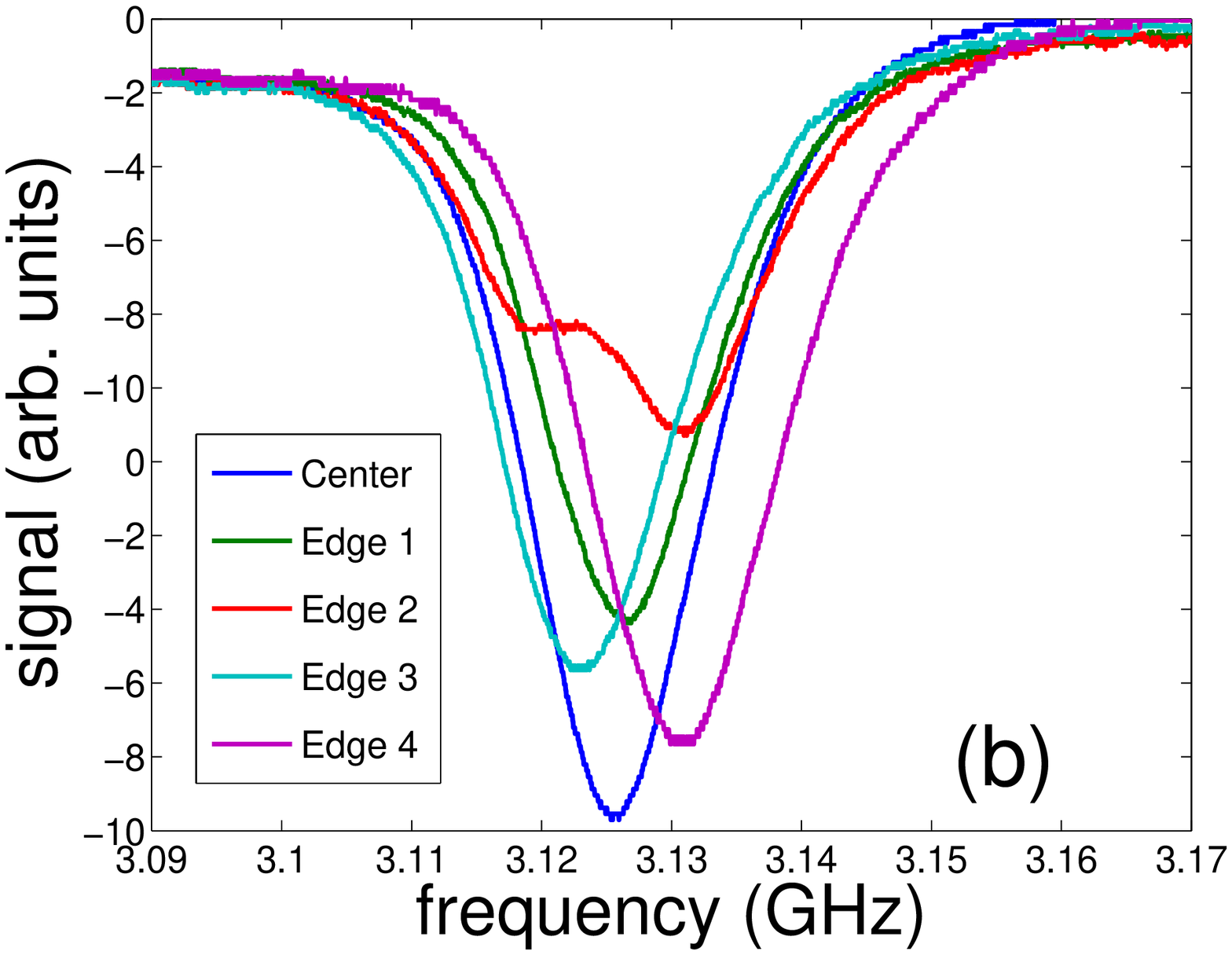} 
\par\end{center}
\end{minipage}
\caption{\label{one} {\small (Color online) (a) Calculated difference in transition frequencies within the ground-state at a magnetic field of 50 mT as a function of the angle between the NV center axis and the external magnetic field with an angle dependence of the transition $m_s$=0$\rightarrow$1 ($m_s$=0$\rightarrow$-1) of 10 (25) MHz/degree. (b) ODMR measurements of the $m_s$=0$\rightarrow$-1 transition of the three non-aligned orientations, at a field of $\sim$ 50 mT. The different curves correspond to the four edges and the center on the diamond sample.}}
\end{minipage}
\end{figure*}
A careful orientation of the crystal with respect to the magnetic field was guided by this premise, and it was physically controlled by slight adjustments in a series of plastic set screws within a custom-built goniometer attached to the sample holder. Figure Supp. 1(a) illustrates the sensitivities of the ground-state transition frequencies at a magnetic field of $\approx$ 50 mT with respect to azimuthal orientation. With these principles guiding the alignment procedure, angular precisions better than 0.5$^{\circ}$ - corresponding to splittings of $\sim$ 20 MHz between the different NV orientations - are easily achieved. To estimate inhomogeneities in the magnetic field, we performed ODMR measurements targeting the four corners plus the center of the diamond sample. Specifically, the magnetic field inhomogeneity is manifested in two aspects: magnitude and orientation. The former is displayed in the frequency distribution of the resonances at various positions on the diamond, while the latter is illustrated for instance by the red curve in Supp. 1(b), corresponding to an ``imperfect alignment''. Judging 
from the residual shifts observed on the resonance lines, it was concluded that an inhomogeneity of the magnetic field on the order of $\sim$ 0.1-0.2 mT in the axial and radial directions, characterized our fields throughout the samples.

\section {II. SAMPLE INFORMATION AND ADDITIONAL RELAXATION MEASUREMENTS}

The samples used (D01 and D02) are synthetic, high pressure, high temperature diamond single-crystals (3 mm*3 mm*0.5 mm), with a natural abundance of $^{13}$C and an initial nitrogen concentration of less than 200 ppm. NV centers were generated through electron irradiation (10 MeV electrons with a dose of $10^{18}$ cm$^{-2}$), followed by subsequent annealing of the sample at 1000 $^{\circ}$C for 2 hours. The comparison of the obtained spin-lattice relaxation times for the samples D01 and D02 showed that both crystals exhibit comparable relaxation characteristics in low field ($\approx$ 50 mT) environments, while at high fields these times differ by ca. an order of magnitude (Supp. 2a). The fluorescence data observed upon laser irradiation suggest that the concentration of NV$^-$ centers in D02 is ca. three times higher than in D01, corresponding to 3 and 10 ppm, respectively. These relaxation and fluorescence data indicate that paramagnetic effects caused by the presence of the NV centers are not the dominant relaxation mechanism for bulk $^{13}$C spins. Interestingly, both samples D01 and D02 exhibit spin-lattice relaxation times at high fields that are several orders of magnitude longer than these at 50 mT (D01: $T_1^{\textit{\tiny LowField}}$ = 2.5 s, $T_1^{\textit{\tiny HighField}}$ = 125 s; D02: $T_1^{\textit{\tiny LowField}}$ = 4.5 s, $T_1^{\textit{\tiny HighField}}\sim$ 1200 s).
\begin{figure*}[h]
\centering{}
\begin{minipage}[t]{17.8cm}
\begin{minipage}[c]{85mm}
\begin{center}
\includegraphics[width=75mm,height=55mm]{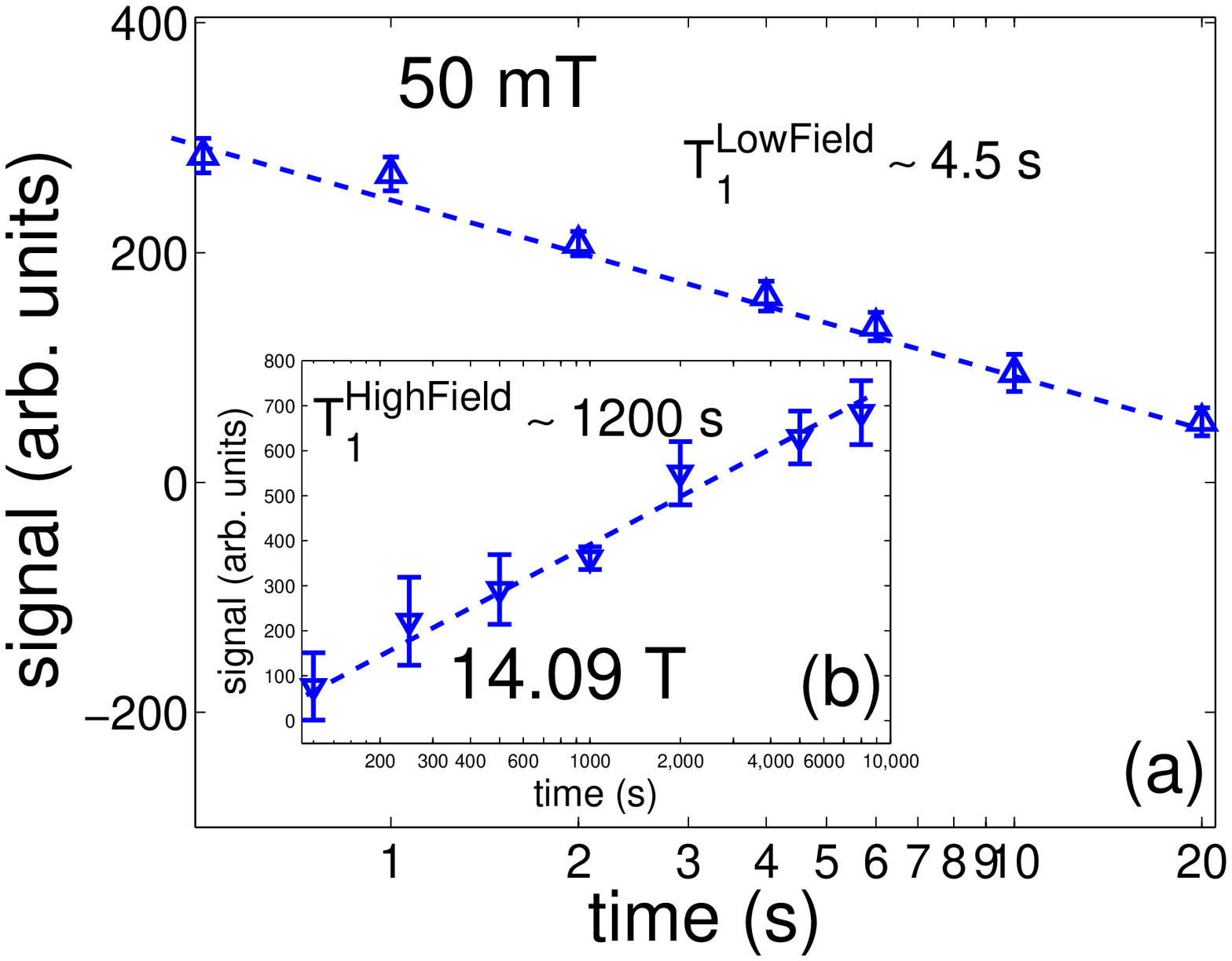} 
\par\end{center}
\end{minipage}
\begin{minipage}[c]{85mm}
\begin{center}
\includegraphics[width=75mm,height=55mm]{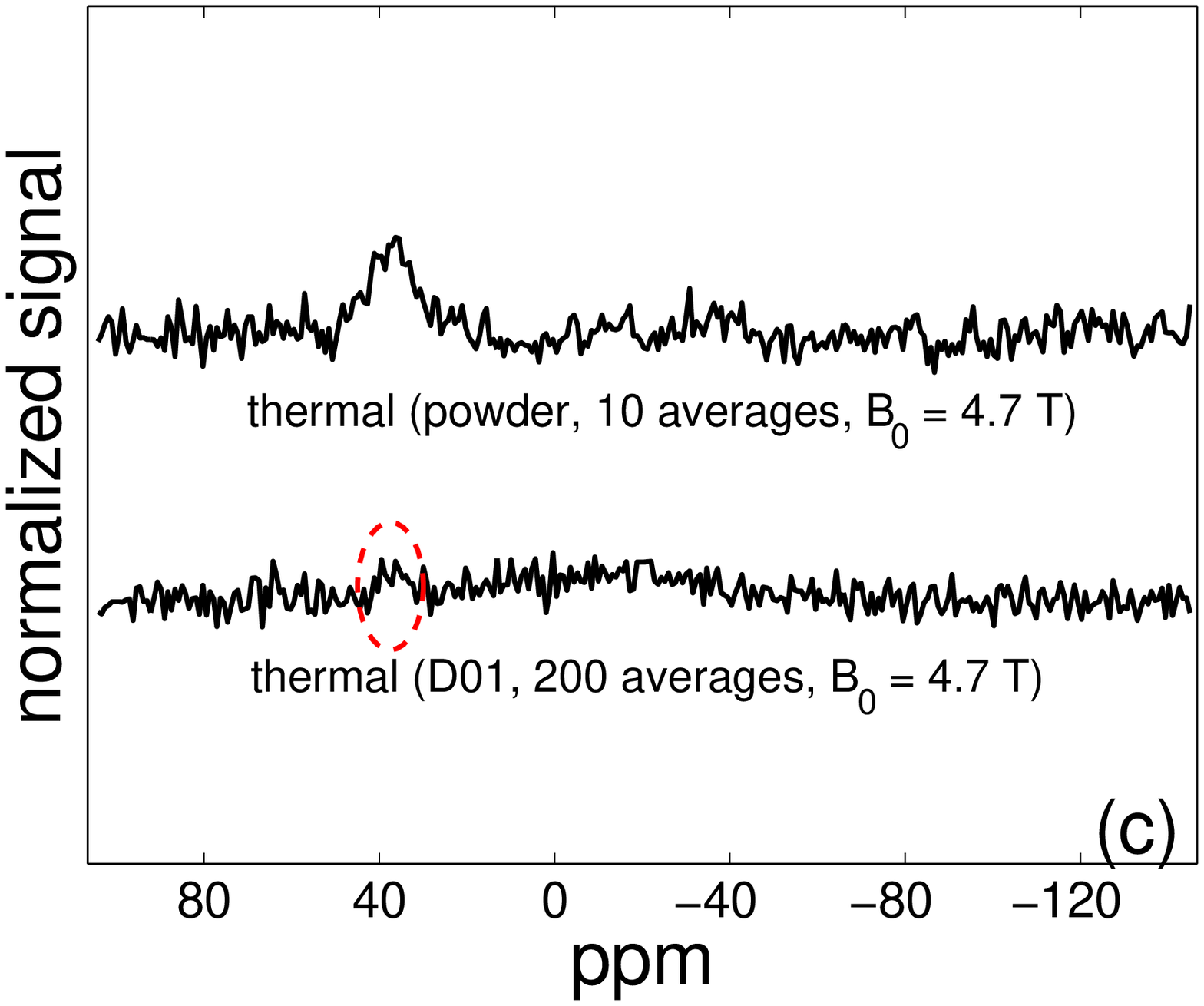} 
\par\end{center}
\end{minipage}
\caption{\label{one} {\small (Color online) (a) Spin-lattice relaxation of optically pumped diamond sample D02 at a polarizing field of ($B_z\sim$ 50 mT, $T_1^{\textit{\tiny LowField}}\sim$ 4.5 s). (b) Relaxation time ($T_1^{\textit{\tiny HighField}}\sim$ 1200 s) obtained for a saturation-recovery experiment under thermal polarizing conditions in a high magnetic field ($B_0$ = 14.09 T). These spin-lattice relaxation times were measured in complete analogy with the methods and equipment used for the diamond crystal D01 (Fig. 3c). (c) Bulk $^{13}$C signal acquired with signal averaging at 4.7 T for the diamond sample D01 and a powder of nanodiamonds. The dashed oval indicates the position of the NMR signal from the single-crystal sample. The powder completely filled the detection coil corresponding to a $\sim$ 50 times greater mass compared to single-crystal samples, leading to a significant reduction of the error margins of the enhancement factors to $\sim$ 10 \%.}}
\end{minipage}
\end{figure*}

\section{III. DENSITY MATRIX SIMULATIONS}

Simulations of the kind illustrated in Fig. 1c of the main text, estimating the transfer of polarization between an electronic and a proximal nuclear spin ensemble  as a function of the magnetic field, were based on calculating the steady-state solution of the spin density matrix based on the master equation

\begin{equation} {d \over dt} \hat{\rho} = {1 \over i\hbar} [\hat{H},\hat{\rho}] + \mathfrak{L}\:\hat{\rho} = 0. \end{equation}
Here $\hat{\rho}$ is the density matrix of the joint manifold of the NV center triplet excited state $^{3}$E and a single proximal $^{13}$C nuclear spin, $\hat{H}$ is the Hamiltonian of the system and $\mathfrak{L}$ is the Lindblad super-operator describing relaxation mechanisms. The Hamiltonian used in the calculation is a given in the paper's Eq. (1) , while the relaxation super-operator can be written as \cite{Shuk08}
\begin{equation} \mathfrak{L}\:\hat{\rho} = \sum_n (\hat{C}_n\:\hat{\rho}\:\hat{C}_n^{\dagger} - {1 \over 2} \hat{C}_n^{\dagger}\:\hat{C}_n\hat{\rho} - {1 \over 2} \hat{\rho}\:\hat{C}_n^{\dagger}\:\hat{C}_n  ), \end{equation}
where $C_n$ are the Lindblad operators which determine the relaxation processes - either decoherence or depopulation. The simulation starts with setting conditions that enabled the establishment of a steady-state polarization, using only the $C_n$ operators. This can be carried out by considering that a given $\hat{C}_n$ = $\gamma_n\:|i\rangle\langle j|$ operator, will result in a net polarization transfer from state $|j\rangle$ to state $|i\rangle$, where $\gamma_n$ are the decay rates associated with the relaxation processes. These ``depopulation operators''
 can therefore account for the spin-lattice relaxation times $T_1$ of the electronic and/or nuclear systems, as well as for the optical pumping process. The latter was modeled by giving different decay rates $\gamma_n$ for the operators $\hat{C}_0$ = $\gamma_0\:|0\rangle\langle \pm 1|$ $\:$ ($\hat{C}_{\pm 1}$ = $\gamma_{\pm 1}\:|\pm 1\rangle\langle 0, \pm 1|$) leading to population transfer from the $m_s$=$\pm$1$\rightarrow$0 $\:$ ($m_s$=0,$\pm$1$\rightarrow \pm$1). The ratio of $\gamma_0$ and $\gamma_{\pm 1}$ was set to 20 to create a steady-state population ($\sim$ 0.9) consistent which experimental data \cite{Man06}. The hyperfine interaction tensor utilized in these simulations was taken from \cite{Chil06}
\begin{equation} A_{IS} = \left( \begin{array}{ccc} 5.0 & -6.3 & -2.9 \\ -6.3 & 4.2 & -2.3 \\ -2.9 & -2.3 & 8.2  \end{array} \right),  \end{equation}
in which all values are given in MHz. From the overall coupling strength of the tensor we can deduce that this $^{13}$C is found three sites from the NV center. In order to mimic different orientations and magnitudes of the nuclear-electronic coupling throughout the diamond, we re-oriented the hyperfine tensor by a spatial rotation using Euler angles ($\alpha$, $\beta$ and $\gamma$). These rotation angles were chosen randomly to create the polarization plots displayed in Figure 1(c) of the main text. We emphasize that this hyperfine tensor is a representation for a proximal $^{13}C$ nuclear spin interacting with an NV center, and allows us to qualitatively reproduce the magnetic field dependence of the hyperpolarized signal. Dependable hyperfine tensors for the full set of $^{13}C$ spins interacting meaningfully with an NV center in the excited-state, are to our knowledge currently not available.

\section{IV. ESTIMATE OF THE $^{13}$C ENHANCEMENT FACTORS}

Two routes were adopted to estimate the $^{13}$C enhancement factors originating from the optical pumping process. One was based on investigating the same single crystals as were pumped, but under thermal equilibrium at high field. Unfortunately, the signal-to-noise ratios of these high-field thermal NMR $^{13}$C experiments were relatively poor (Fig. 3a of the main text); in addition to the relatively low gyromagnetic factor and natural abundance of $^{13}$C this was a consequence of a filling factor of the coil that needed to be compatible with sample shuttling. Given the long high-field nuclear spin-lattice relaxation times in diamond ($T_1 \sim$ 120 s), this made the estimate of the enhancement factors prone to statistical errors, despite prolonged averaging. Still these measurements lead to estimate enhancement factors on the order of 125 (500) for the diamond crystal D01 (D02), with about 50\% uncertainty in each case. 
An alternative route was also taken to estimate the optically-derived enhancement factor, based on the use of a nanodiamond powder. The nanodiamonds exhibited much shorter $^{13}$C longitudinal relaxation times ($T_1\sim$ 300 ms) and an increased SNR due to the improved filling factor of the detection coil (Supp. 2c). A more accurate enhancement factor $\eta$ was determined from these high-field spectra as 
\begin{equation} \eta =  \frac{S_{\textit{OP}}\:\cdot\:\sqrt{nt_{\textit{powder}}}\:\cdot\:m_{\textit{powder}}}{S_{\textit{powder}}\:\cdot\:\sqrt{nt_{\textit{OP}}}\:\cdot\:m_{\textit{OP}}}, \end{equation}    
where $m$ accounts for the different masses of the optically-pumped and thermal powder samples ($m_{\textit{powder}}$ = 800 mg, $m_{\textit{OP}}$ = 16 mg), \textsl{nt} are the number of averaged transients ($\textit{nt}_{\textit{powder}}$ = 10, $\textit{nt}_{\textit{OP}}$ = 1), and $S$ are the integrated $^{13}$C spectral signal intensities. The error of $\eta$ is solely determined by the {uncertainties of these intensities. The enhancement factor for the diamond sample D02 after 10 s of irradiation at 7 W, could be thus estimated as 
\begin{equation}\nonumber \eta = 486 \pm 51; \end{equation} 
in agreement with the single-crystal derivation.

\section {V. ADDITIONAL EXPERIMENTAL CONSIDERATIONS ON THE DEFINITION OF ``BULK'' NMR SIGNALS}}

The relatively high concentration of paramagnetic impurities (NV $\sim$ 10 ppm, substitutional nitrogen $\lesssim$ 200 ppm) in the used diamond crystals, makes an accurate definition of bulk ensemble challenging. In such natural-abundance $^{13}$C diamond samples the electron-nucleus coupling is found to be always orders of magnitude stronger than the dipole-dipole interaction between the nuclei. Therefore, the nuclear spin-lattice ($T_1$) and spin-spin ($T_2$) relaxation times are governed by the field fluctuations of the electronic defects, and both the linewidth and chemical shift characteristics of the corresponding nuclear polarizations are greatly influenced by the type of the paramagnetic impurities and their concentration. Our experiments show that the NMR signals obtained through optical pumping and shuttling $^{13}$C measurements, exhibit the same chemical shift, linewidth and relaxation characteristics as the NMR signal obtained at high field via conventional signal averaging. In an effort to further characterize the nature of the optically pumped $^{13}$C ensemble that is observed in the shuttling experiments, these $^{13}$C measurements were complemented by the low-field nuclear manipulations referred to in the ``Results'' section. 
These measurements probed the $^{13}$C environment via resonant radiofrequency pulses at frequency $\gamma_{^{13}C}\cdot B_z\sim$ 545 kHz, applied at the ESLAC conditions and generated by an additional Helmholtz coil (15 coil windings, 5 mm diameter) oriented perpendicularly to the axial magnetic field $B_z$. A complete inversion of the nuclear macroscopic polarization (nuclear 180$^{\circ}$ pulse) was achieved by applying an RF pulse with a duration of $t_{\textit{180}}\sim$ 90 $\mu$s (Supp. 3a) with an irradiation power of $\sim$ 150 W, corresponding to a magnetic field of $\approx$ 5 G. By applying these narrow-band low-field RF pulses at different stages of the pump-shuttle process, additional information about the spatial distribution of the nuclear spins contributing to the optically-enhanced NMR signal could be obtained. NMR signals obtained in two shuttling experiments were compared: in one of these a low-field RF 180$^{\circ}$ pulse was applied during optical pumping, before laser illumination was stopped; in a second experiment an identical 180$^{\circ}$ pulse was applied after the laser illumination was stopped (Supp. 3b). The post-illumination delay $\tau$ was chosen long (> 50 ms) in comparison to the electronic spin-lattice relaxation  time, thereby ensuring that the optically pumped electronic system, which mainly involved $m_S$ = 0 in the first of the experiments, had returned to a fully relaxed set of nearly equally populated electronic states in the second experiment. If the signal were to originate predominantly from $^{13}$C spins exhibiting a strong hyperfine interaction ($\frac{A_{IS}}{2\pi}\gg t^{-1}_{\textit{180}}$) the RF pulse applied in the latter experiment would only affect the fraction of $^{13}$C spins coupled to an electron in the $m_S$=0 state, resulting in a three times smaller signal. If on the other hand the majority of signal had originated from ``bulk'' 
$^{13}$C spins exhibiting small electron-nuclei couplings ($\frac{A_{IS}}{2\pi}\le t^{-1}_{\textit{180}}$), the NMR signal of both experiments should agree within experimental errors. As evidenced by the results in Supp. 3(c), the latter scenario is the one actually realized. The increasing deviation for longer time delays |$\tau$| can be interpreted by repolarization effects imparted by the optical pumping as the inversion of the nuclear polarization departs from the $\tau$ = 0 instant. 
Thus, it can be concluded that the majority of the nuclear spins contributing to the NMR signal exhibit small hyperfine interactions ($\frac{A_{IS}}{2\pi} < t^{-1}_{\textit{180}}\sim$ 10 kHz) corresponding to distances of 1 nm or more from paramagnetic defects.
\begin{figure*}[h]
\centering{}
\begin{minipage}[t]{17.8cm}
\begin{minipage}[c]{85mm}
\begin{center}
\includegraphics[width=75mm,height=55mm]{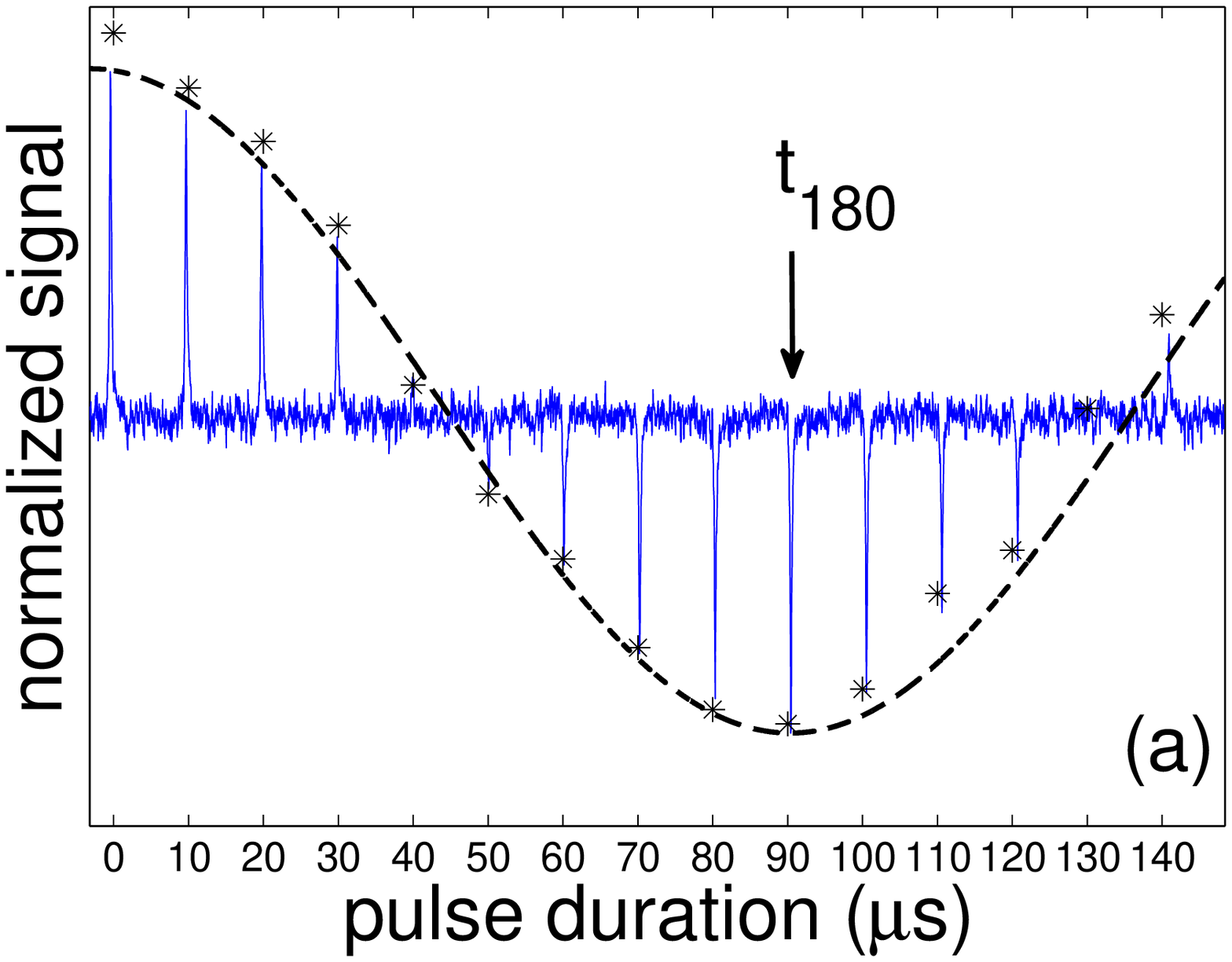} 
\par\end{center}
\end{minipage}
\begin{minipage}[c]{85mm}
\begin{center}
\includegraphics[width=75mm,height=55mm]{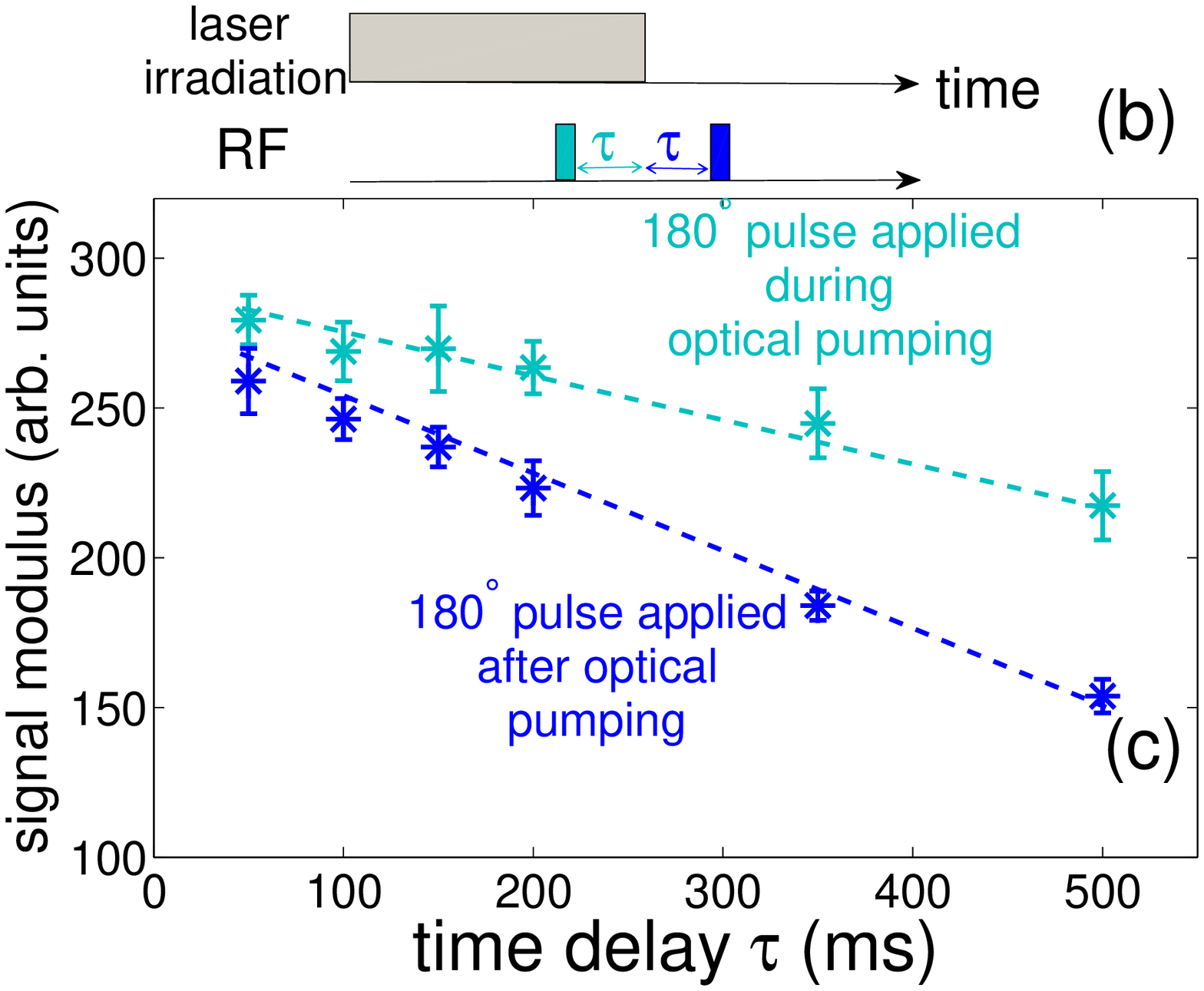} 
\par\end{center}
\end{minipage}
\caption{\label{one} {\small (Color online) (a) Experimentally observed $^{13}$C nutation curve used for the calibration of the tip angle of the resonant RF pulses ($\omega$ = 545 kHz) applied at 50 mT. Single-shot, high-field NMR spectra (blue) are observed after 10 s of an optical pumping/sample shuttling procedure for low-field RF pulses applied with variable pulse durations (steps of 10 $\mu$s) and spliced together and illustrated in combination with the corresponding integrated spectral signal (black dashed line, asterisks).
A complete inversion of the nuclear polarization ($\pi$-pulse) is found for an RF power of $\sim$ 150 W, corresponding to a magnetic field of $\approx$ 5 G and a duration of $\sim$ 90$\mu$s. (b-c) Comparison of the observed bulk $^{13}$C NMR signals arising from the optically pumped sample D02, in experiments where the  polarization was inverted via a resonant 180$^{\circ}$-pulse ($\gamma_{^{13}C}\cdot B_z$ = 545 kHz, $t_{\textit{180}}\sim$ 90 $\mu$s) before (cyan) and after (blue) the end of the optical pumping for different time delays $\tau$.}}
\end{minipage}
\end{figure*}
This low threshold on the hyperfine interaction strengths characterizing the nuclei contributing to the observed NMR signal, further corroborates that the nuclear polarization in both the optically-pumped and thermally-equilibrated scenarios originate from the same, weakly coupled spin ensemble. 

\section{VI. ON THE NUMBER AND EXTENT OF THE $^{13}$C SPINS POLARIZED PER NV CENTER}

In the following we estimate the number of polarized $^{13}$C's per NV center, and briefly discuss how the spatial propagation of the electron-nuclear polarization could proceed away from the NV centers - either by spin diffusion or by a ``direct'' 
polarization process. The ratio of polarized $^{13}$C to NV centers is estimated through the following
\begin{equation}\label{ratio} {{c_{^{13}C}\cdot P_{^{13}C}}\over c_{NV}} = {0.011\cdot 0.005\over {1\over 4}\cdot 10\cdot 10^{-6}} \approx 20.\end{equation}
Here $c_{^{13}C}$ is the natural abundance of $^{13}$C spins, $P_{^{13}C}$ is their polarization, and $c_{NV}$ is the concentration of NV centers aligned to the magnetic field. 
To assess the nature of the electron $\leftrightarrow$ nucleus polarization transfer, we employ a simple model whereby statistically distributed NV centers in the sample create ``spheres''
containing multiple $^{13}$C nuclei that get polarized by individual NV centers. This model is based on diamonds containing NV center concentrations in the order of $c_{NV}\sim$ 10 ppm, which is the estimated NV center concentration in the sample D02, as judged by comparing the fluorescence signal of our samples with the reports given by Acosta et al. \cite{Acos09}. For this NV center concentration the average distance $d_{\textit{avg}}$ between defects can be calculated as
\begin{equation}\label{eq6} d_{\textit{avg}} = \left( N_A\cdot {\rho_{\textit{dia}}\over M_{\textit{dia}}}\cdot c_{NV} \right)^{-\frac{1}{3}} = \left( 6.023\cdot 10^{23}\:{1\over mol}\cdot {3.52\:{g\over cm^{3}}\over 12.01\:{g\over mol}}\cdot 10\cdot10^{-6} \right)^{-\frac{1}{3}} \approx 8.3\:\textit{nm}. \end{equation}
Here $N_A$ is Avogadro's number, and $\rho_{\textit{dia}}$ and $M_{\textit{dia}}$ are density and molar mass. As mentioned above, only one out of four possible crystallographic directions of the NV centers in a single crystal (e.g. the [111] one) contributes to the electron$\leftrightarrow$nuclei polarization transfer process. This corresponds to only 25 \% 
of the diamond's NV centers, increasing the average distance between polarizing centers to 
\begin{equation}\nonumber d_{\textit{avg}} \sim 13 \:\textit{nm}. \end{equation}

\subsection{VI-I. SPIN DIFFUSION}

Measurements by Dr\'{e}au 
and co-workers \cite{Drea12} of optically detectable nuclear spins indicated that hyperfine interactions of 0.2-1 MHz may lead to significant nuclear polarizations (more than 10\%) 
via optical pumping. Taken such conditions as defining the minimal demands for a meaningful direct polarization, it follows that the direct polarization of ``bulk'' nuclear 
spins with smaller electron-nucleus couplings will be negligible. A spin-diffusive transfer mechanism propagating nuclear polarization from the strongly coupled $^{13}$C's to the bulk could then account for the latter's polarization. In such scenarios the laser irradiation of the NV defects will lead to highly polarized electronic states, in which $\sim$ 85\% 
of the electronic spins are found in the $m_S$ = 0 state. The absence of a strong hyperfine interaction with the surrounding nuclei should enable a much weaker net of dipole-dipole couplings arising between $^{13}$C spins in the sample, to drive a carbon-carbon spin diffusion-type process via $\hat{I}_{1+}\hat{I}_{2-}+\hat{I}_{1-}\hat{I}_{2+}$-like terms. This serves to transport out the electron's optically enhanced polarization towards nuclear spin species detectable by means of NMR. To estimate the approximate number of $^{13}$C spins that could be polarized in this manner, we approximate the characteristic length as
\begin{equation}r_{\textit{pol}} = \sqrt{D\cdot \tau}, \end{equation}
where $D$ = 6.7$\cdot 10^{-15}\:{cm^2\over s}$ is the spin diffusion coefficient calculated for diamond with a natural abundance of $^{13}$C \cite{Corn01} and $\tau$ is the time constant defining the establishment of $^{13}$C polarization. For pumping times of $\tau$ = 10 s like those supported by the analyzed crystals (Fig. 3d), the characteristic diffusion length is $r_{\textit{pol}}\approx$ 2.5 nm, corresponding to a hyperfine interaction of $\frac{A_{IS}}{2\pi}\sim$ 1.3 kHz. Together with the minimum distance between NV defects and $^{13}$C spins established in low-field RF pulse experiments (Supp. 3b) it can be concluded that only $^{13}$C's within a small shell (1 nm $\le$ r $\le$ 2.5 nm) of the initial polarization spheres contribute to the observed NMR signal. It is worthwhile noting that the spin diffusion coefficient (and therefore the characteristic length $r_{\textit{pol}}$) just used was determined for a face-centered cubic lattice system, neglecting the influence of the relatively high concentration of paramagnetic defects. Dependable experimental findings about the characteristics of the nuclear spin diffusion process as a function of the concentration of paramagnetic defects are, to our knowledge, not available. Consequently, the reported diffusion lengths of 2.5 nm should be interpreted as lower bounds of the spin diffusion process. 

Following the laser irradiation of the crystal defects and the corresponding distribution of polarization via spin diffusion, our experiment proceeds by shuttling the sample from the polarizing field to a 4.7 T detection field. The observed NMR signal consisted of a single optically-enhanced $^{13}$C spectral line with a linewidth of $\le$ 1 kHz. While potentially present, no satellite signals originating from hyperfine triplet splittings were observed; apparently these peaks were too weak to be distinguished from the background noise. To better estimate the degree of nuclear polarization underlying this signal it is worth noting that the electronic spin-lattice relaxation time was about two orders of magnitude shorter than the timescale demanded by the sample shuttling process; upon arriving to the NMR detection region, the electronic spins were thus fully-relaxed, with nearly equal populations of the NV electronic energy levels. It follows that for polarized $^{13}$C to contribute to a linewidth like the one observed by NMR, they either (i) possessed a hyperfine interaction to the NV smaller than the detected linewidth of 1 kHz; or (ii) they had large hyperfine couplings, but only reflect the NV center's $m_S$ = 0 state. As mentioned in the preceding paragraph, the largest NV-nucleus distances that are compatible with the experimental build-up times $\tau$ correspond to a characteristic length  $r_{\textit{pol}} \approx$ 2.5 nm, implying effective nuclear-electron hyperfine couplings of $\ge$ 1.3 kHz. Consequently, the observed $^{13}$C line shapes suggest that only a fraction of the nuclear spins residing in the aforementioned ``polarization spheres'' 
will contribute to the NMR signal, namely those coupled to NV defects that are predominately in the $m_S$ = 0 state. This in turn suggests that the measured enhancement $^{13}$C factors further underestimate the real nuclear polarization achieved by optical pumping. It is also suggesting that the local polarization within these polarization spheres can be considerably higher, reaching possibly tens of percents.

\subsection{VI-II. DIRECT POLARIZATION}

In section VI-I we assumed a model whereby strongly hyperfine coupled nuclei \cite{Drea12} undergo a ``direct'' polarization 
transfer process, which subsequently spread by spin-diffusion. The occurrence of such a process follows the experimental data arising from ODMR spectroscopy \cite{Man06}. It is conceivable, however, that ``weak'' hyperfine 
couplings ($A_{IS}\le$ 200 kHz) can also induce significant polarization transfers, provided that the laser illumination is chosen long enough ($\tau$ $\gg$ excited state life-time) to permit a large number of electronic excitation/relaxation cycles. Unlike the relatively simple diffusion model used in section VI-I, the magnitude and time scale of such ``direct''
polarization process would depend on the direction and strength of the hyperfine coupling, the electronic pumping rate, the excited-state life-time and the relaxation rates of the involved electronic and nuclear states. In analogy to supplementary section II, we employed a toy-model density matrix simulation, to estimate the approximate amount of polarization transferred to a single nucleus as a function of the strength of the hyperfine interaction. The simulations (Supp. 4) suggest that hyperfine interactions as weak as $\sim$ 1 kHz (corresponding to a NV-nucleus distance of $\sim$ 2.5 nm) can polarize $^{13}$C spins on the time scale of their spin-lattice relaxation. Thus, we conclude that direct polarization driven by interaction of a single NV center and a single nuclear spin can potentially compete and/or complement the spin diffusion process described earlier. 
\begin{figure*}[h]
\centering{}
\begin{center}
\includegraphics[width=75mm,height=55mm]{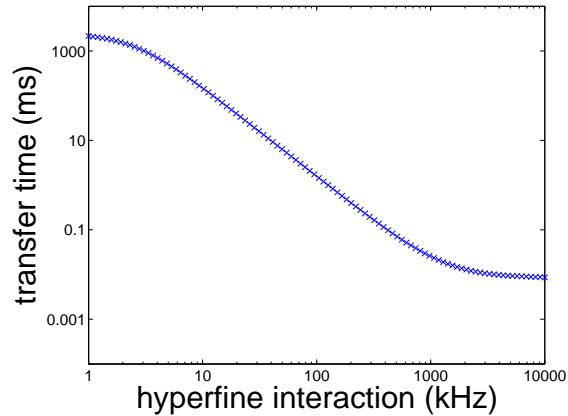} 
\par\end{center}
\caption {\label{supple4} {\small (Color online) Nuclear polarization time scales as a function of the hyperfine interaction, determined by a toy-model density matrix simulations of a single NV center coupled to a single $^{13}$C spin (as introduced in section II).}}
\end{figure*}
A more complete and reliable picture of the polarization transfer process should include both direct electron-nuclear and spin-diffusion-like nuclear-nuclear coupling terms. The Hamiltonian of a single NV center interacting with an ensemble of nuclear spins should therefore be of the form
\begin{equation}\label{eq8} H_{complete} = H_{NV} + H_{Zeeman} + \sum_i \hat{S}\cdot A_{IS}^i\cdot \hat{I_i} + \sum_{i,j} \hat{I_i}\cdot A_{I_1I_2}^{ij}\cdot \hat{I_j}. \end{equation}
An analysis of this Hamiltonian could provide interesting insights into the dynamics of both processes. 
%Further research into how to assess this kind of problem is in order.